\documentclass[aps,pre, preprint,superscriptaddress]{revtex4}
\usepackage{graphicx}
\usepackage{epstopdf}
\usepackage{amsmath,amssymb}
\usepackage{float}

\newcommand{\ignore}[1]{}
\newcommand{\beq}{\begin{equation}}
\newcommand{\eeq}{\end{equation}}
\begin{document}

\title
{Multistability in an unusual phase diagram induced by the competition between 
antiferromagnetic-like short-range and ferromagnetic-like long-range interactions}

\author{Masamichi Nishino}
\email[Corresponding author: ]{nishino.masamichi@nims.go.jp} 

\affiliation{Research Center for Advanced Measurement and Characterization, National Institute for Materials Science, Tsukuba, Ibaraki 305-0047, Japan}

\affiliation{International Center for Materials Nanoarchitectonics, National Institute for Materials Science, Tsukuba, Ibaraki 305-0044, Japan} 

\author{Per Arne Rikvold}
\affiliation{Department of Physics, Florida State University, 
Tallahassee, FL 32306-4350, USA}
\affiliation{PoreLab, Department of Physics, University of Oslo, 
P.\ O.\ Box 1048, Blindern, 0316 Oslo, Norway}
\altaffiliation[]{Current address.}

\author{Conor Omand} 
\affiliation{Department of Physics, Graduate School of Science, The University of Tokyo, Bunkyo-Ku, Tokyo 113-0033, Japan}

\author{Seiji Miyashita} 
\affiliation{Department of Physics, Graduate School of Science, The University of Tokyo, Bunkyo-Ku, Tokyo 113-0033, Japan}

\date{\today}

\begin{abstract}

The interplay between competing short-range (SR) and long-range (LR) interactions can cause nontrivial structures in phase diagrams. 
Recently, horn-shaped unusual structures were found by Monte Carlo simulations 
in the phase diagram of 
the Ising antiferromagnet (IA) with infinite-range ferromagnetic-like (F) interactions [Phys. Rev. B {\bf 93}, 064109 (2016); {\bf 96}, 174428 (2017)], and also in an IA with LR interactions of elastic origin modeling spin-crossover materials [Phys. Rev. B {\bf 96}, 144425 (2017)]. 
To clarify the nature of the phases associated with the horn structures, we study the phase diagram of the IA model with infinite-range F interactions by applying a variational free energy in a cluster mean-field (CMF) approximation. 
While the simple Bragg-Williams mean-field theory for each sublattice does not produce a horn structure, we find such structures with the CMF method. This confirms that the 
local thermal fluctuations enabled by the multisite clusters are essential for this phenomenon. 
We investigate in detail the structure of metastable phases in the phase diagram.
In contrast to the phase diagram obtained by the Monte Carlo studies, we find a triple point, at which ferromagnetic-like, antiferromagnetic-like, and disordered phases coexist, and also six tristable regions accompanying the horn structure. We also point out that several characteristic endpoints of first-order transitions appear in the phase diagram. 
We propose three possible scenarios for the transitions related to the tristable regions. 
Finally, we discuss the relation between the triple point in this phase diagram and that of a possible lattice-gas model, in which solid, liquid, and gas phases can coexist.
\end{abstract}

%\pacs{75.30.Wx 64.60.-i 75.60.-d 64.60.De}
%\pacs{75.30.Wx 75.50.Xx 75.60.-d 64.60.-i}

\maketitle

%----------------------------------------------------------------------------

\section{Introduction}
\label{sec:I}

The interplay between competing short-range (SR) and long-range (LR) interactions causes complex orderings in many physical systems. 
Recently, an unusual ``horn structure," which is surrounded by ferromagnetic-like (F) spinodal lines, disorder (D) spinodal lines, and a critical line, was found in the phase diagram of the Ising antiferromagnet (IA) with infinite-range F interactions~\cite{Per1,Per3}. A similar horn structure was found in an elastic-interaction model with antiferromagnetic-like (AF) SR interactions, modeling spin-crossover (SC) materials~\cite{Nishino_AF2}. This suggests that such unusual structures are universal in models with competing SR and LR interactions and may be realized in real (experimental) systems including SC materials. 

SC materials show colorful ordered structures and switching phenomena induced by temperature change, pressure variation, light irradiation, etc.~\cite{Gutlich_book,SC_book2,Kahn,Garcia,Bousseksou,Bousseksou2,Pillet1,Tanasa,Pillet2,Brefuel,Chong,Bousseksou3,Slimani,Collet,Slimani2,Felix,Collet2,Bertoni,Watanabe,Rat,Mikolasek}.
In these materials, the SR interactions and the LR interactions of elastic origin compete. 
SC materials have attracted much attention due to their potential applications to memory devices, sensors, etc. 
It has been pointed out that elastic interactions play an important role in cooperativity for SC materials,  
and studies with microscopic elastic interaction models have been performed \cite{Nishino_elastic,Miya,Nicolazzi1,Enachescu1,Nishino_elastic2,Enachescu2,Nicolazzi2,Nakada2,Slimani3,Nishino_AF1,Enachescu3,Kamel_elastic1,Kamel_elastic2,Nishino_triangular,Enachescu4,Nishino_AF2,Mikolasek2}. 

The difference of molecular sizes between the high spin (HS) and low spin (LS) states 
that characterize SC materials causes local lattice distortions, which lead to effective LR elastic interactions.      
A variety of orderings originate from the interplay between direct SR and effective LR interactions of elastic origin. 
The LR interaction induced by lattice elasticity is important in the one-step F-like transition between the LS and HS phases with a second-order (continuous) or first-order (discontinuous) transition. 
The mean-field universality class is realized in the second-order transition~\cite{Miya,Nakada2,Nishino_AF1}. 

Some SC materials exhibit two-step phase transitions~\cite{Koppen,Petrouleas,Jakobi,Real,Boinnard,Buron,Bousseksou4,Shatruk}. 
The elastic interaction model with  AF SR interactions enables us to classify various types of two-step SC transitions between F uniform HS or LS phases and AF checkerboard phases, in which a second-order or first-order transition occurs in each step~\cite{Nishino_AF1,Nishino_AF2}. 
Unlike in the one-step F transition between the LS and HS phases, the SR interactions are essential in second-order (continuous) transitions between the AF and F phases. 
The Ising universality class is realized in these transitions. In contrast, the LR interaction 
is significant in first-order (discontinuous) transitions between the AF and F phases. A new type of two-step SC transition is realized if the horn structures appear~\cite{Nishino_AF2}. 
 
The simplification of the LR interaction obtained by replacing the elastic interaction with infinite-range F interactions causes a qualitatively similar cooperative nature of the bulk properties~\cite{Nishino_triangular,Per1,Per3}. The IA model with infinite-range F interactions is 
therefore better suited for clarifying detailed features of phase diagrams. 
The usual Bragg-Williams (BW) mean-field (MF) theory for each sublattice does not produce such unusual horn structures, even when the infinite-range F interaction is relatively strong \cite{Per1}. 
However, Monte Carlo (MC) methods produce such structures~\cite{Per1,Per3}. 
This suggests that thermal fluctuations are essential for the generation of the unusual structures. 
The infinite-range F interaction is essentially MF in nature, and thus the F order is quite robust against thermal fluctuations.
In contrast, the AF Ising interaction is of short range, and we expect that the ordering caused by these interactions should be strongly affected by thermal fluctuations.  

In the MC studies the ``horn region'' is identified as a region surrounded by F spinodal lines, D spinodal lines, and a critical line~\cite{Nishino_AF2,Per1,Per3}. 
The critical points were determined by the Binder fourth-order cumulant method~\cite{Binder}. 
However, larger error bars for the locations of the crossing points in the Binder plots for different system sizes were observed in the higher field region~\cite{Per3}, and it is difficult to identify if they 
indicate second-order transitions or more complex phase relations.

In order to understand the mechanisms underlying the generation of unusual phase structures, including horn structures, it is important to study how such structures appear as 
thermal fluctuations are introduced into the system. 
In the present study, we therefore investigate the phase diagram for the IA model with 
infinite-range F interactions by a kind of cluster MF (CMF) theory 
\cite{KIKU51, OGUC55,RIKV91,MORI94,ETXE04,PELI05,Agra,YAMA09}, 
which takes into account the SR fluctuations within a finite, multisite cluster. 
The AF ordering requires the use of two sublattices \cite{OGUC55,Agra}, and the structure of the phase diagram is determined by 
evaluating the free-energy landscape of the model by the variational principle 
\cite{KIKU51,MORI94,PELI05,Agra,Fynman-texbook,Callen-textbook}.
%and solving the variational equations for the free energy.
The resulting phase diagrams contain various metastable phases. 
In particular, we find six regions in which one of the three phases, AF, F, or D, is globally 
stable, and the other two are metastable.  
Such ``tristable" regions were not found  in the MC studies. 
We discuss the characteristic features of the multistability of the metastable phases and present possible new scenarios for the associated phase transitions.

The rest of this paper is organized as follows. In Sec.\ \ref{sec_model} the
model and method are presented. The CMF theory is developed, and 
the free energy and its variational equations are derived. 
%The details of the derivation are given in appendix A.  
Section \ref{sec_phase_diag} is devoted to the results and discussion. 
The details of the phase diagram are shown, focusing on the multistability. 
In Sec.\ \ref{summary} we give discussion and summary. 
The distinctions between the variational parameters of the 
variational MF method and the order parameters of the system are discussed in the Appendix.

\section{Model and method}
\label{sec_model} 

\subsection{Model}

\begin{figure}
\centerline{
\includegraphics[clip,width=5.0cm]{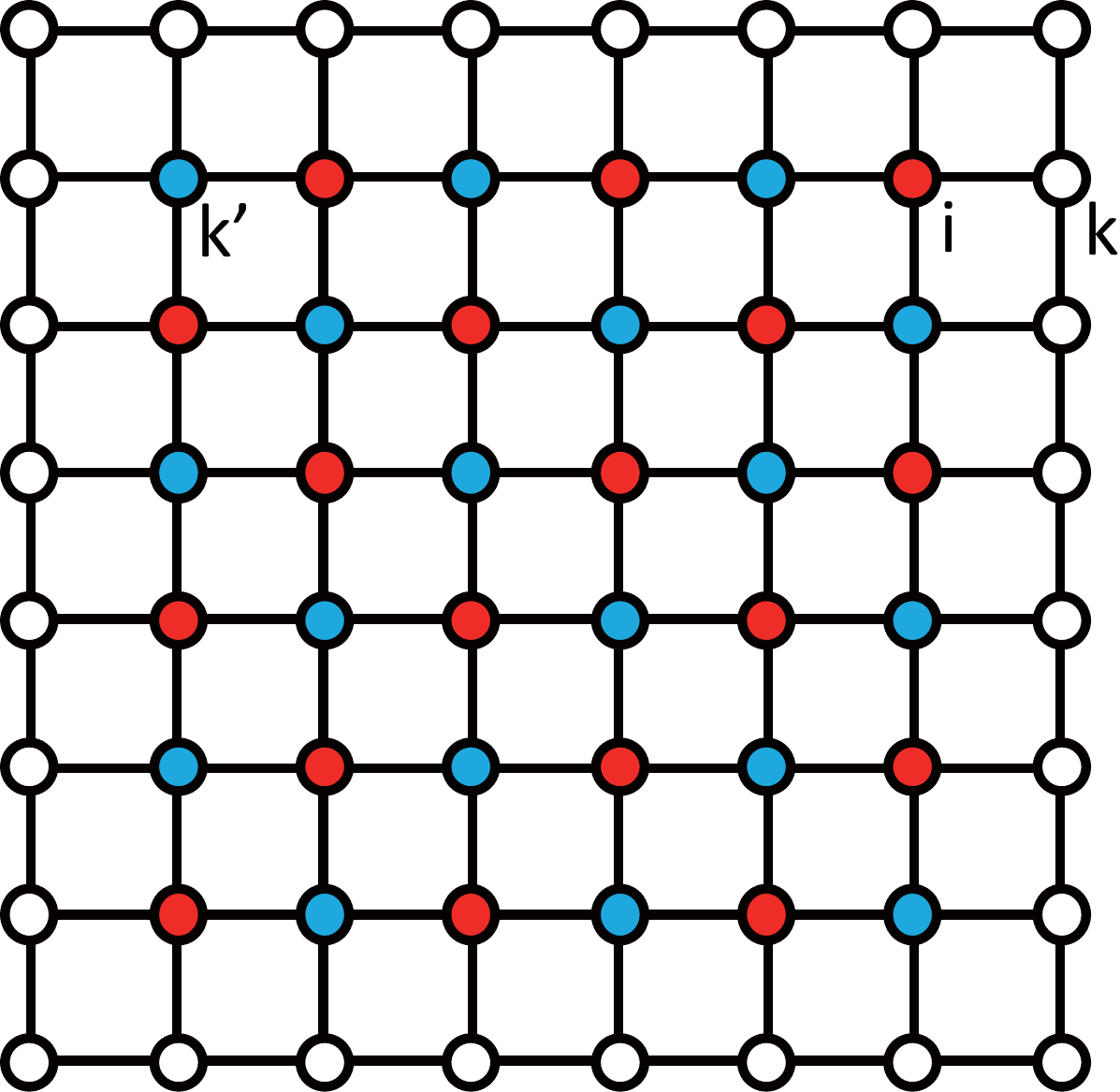}} 
\caption{A cluster which consists of $L \times L$ sites (blue and red filled circles) 
embedded in a system with $N_{\rm C}$ clusters. 
Open circles denote neighboring sites in neighboring clusters. 
Each site in the cluster belongs to sublattice A (blue) or B (red) in the bipartite lattice of the system.
The indices $i$ and $k$ are used in Eqs. (8-27). Site $k'$ in a cluster is equivalent to site $k$ in a neighboring cluster. }
\label{Fig1_Cluster}
\end{figure}

We study a model which consists of SR and LR interactions, 
\beq
{\cal H}={\cal H}_1+{\cal H}_2, 
\label{Ham}
\eeq
where ${\cal H}_1$ is the nearest-neighbor $S=1/2$ Ising antiferromagnet on a square lattice, 
\begin{align}
{\cal H}_1= J \sum_{\langle i,j \rangle} \sigma_i \sigma_j \;.
\label{Ham1}
\end{align}
Here, $\sigma_i=\pm 1$, $J>0$ induces a staggered order, and $\langle i,j \rangle $ denotes summation over nearest-neighbor pairs. 
${\cal H}_2$ gives the infinite-range F and Zeeman interactions, 
\begin{align}
{\cal H}_2= -\frac{A}{2N}  \Big( \sum_{i=1}^N  \sigma_i    \Big)^2
-H \sum_{i=1}^N \sigma_i. 
\label{Ham2}
\end{align}

\subsection{Free energy}

We construct a variational free energy $F_{\rm v}$ by applying the Bogoliubov inequality~\cite{Agra,Fynman-texbook,Callen-textbook}, 
\beq
F \le F_{\rm v}=F_{\rm CMF} + \langle {\cal H}-{\cal H}_{\rm CMF} \rangle.
\eeq
Here ${\cal H}$ is the exact Hamiltonian and ${\cal H}_{\rm CMF}$ is the cluster mean-field (CMF) Hamiltonian defined below. The CMF free energy is defined as 
 \beq
F_{\rm CMF} = -\frac{1}{\beta} \ln Z_{\rm CMF},  
\eeq
where 
$Z_{\rm CMF}=\mathrm{Tr} \mathrm{e}^{-\beta {\cal H}_{\rm CMF}}$, and 
\beq 
\langle X \rangle \equiv \frac{1}{Z_{\rm CMF}} \mathrm{Tr} X \mathrm{e}^{-\beta {\cal H}_{\rm CMF}}.
\eeq 
We emphasize that the statistical average is taken with ${\cal H}_{\rm CMF}$ instead of ${\cal H}$. 
Here $\beta$ is the inverse temperature: $\beta=\frac{1}{k_{\rm B}T}$.

The BW MF approximation fails to reproduce the horn structure, which requires  
the effects of SR fluctuations to be included into the model. 
For this purpose, we divide the lattice into $N_{\rm C}$ equivalent clusters, each of which has a size of $L \times L$ (see Fig.~\ref{Fig1_Cluster}).  
The total system size is $N=N_{\rm C} \times L \times L$. 
Then, ${\cal H}_1$ is rewritten as 
\begin{align}
{\cal H}_1= \sum_{i_{\rm C}=1}^{N_{\rm C}} J \Big(\sum_{\langle i,j \rangle_{i_{\rm C}}} \sigma_i \sigma_j + \frac{1}{2} \sum_{\langle \langle i,k \rangle \rangle_{i_{\rm C}}}^{4L} \sigma_i \sigma_k \Big). 
\label{H1}
\end{align}
Here $\langle i,j \rangle_{i_{\rm C}}$ denotes a nearest-neighbor pair with both sites 
in the ${i_{\rm C}}$th cluster, and $\langle \langle i,k \rangle \rangle_{i_{\rm C}}$ a nearest-neighbor pair at the border of the cluster, i.e., site $i$ belongs to the ${i_{\rm C}}$th cluster and site $k$ to a neighboring cluster.

We focus on the interactions at the border between two clusters, e.g., the interaction between sites $i$ and $k$ in Fig.~\ref{Fig1_Cluster}, which is approximately given by 
\begin{align}
\sigma_i \sigma_k \simeq \sigma_i \langle \sigma_k \rangle + \langle \sigma_i \rangle \sigma_k -\langle \sigma_i \rangle \langle \sigma_k \rangle. 
\end{align} 
Here we replace the average of border site $\langle \sigma_k \rangle$  by variational parameter $x$ ($y$) on sublattice A (B). 
The same replacement is performed to $\langle \sigma_i \rangle$. Thus, for sites $i$ and $k$ on sublattices B and A, respectively, 
\begin{align}
\sigma_i \sigma_k \simeq \sigma_i x + y \sigma_k-xy, 
\end{align}
and for sites $i$ and $k$ on sublattices A and B, respectively,
\begin{align}
\sigma_i \sigma_k \simeq \sigma_i y + x \sigma_k-xy. 
\end{align}

Then we construct a CMF Hamiltonian for ${\cal H}_1$ as follows, 
\begin{align}
{\cal H}_{\rm 1, CMF}(x,y) &=J \sum_{i_{\rm C}=1}^{N_{\rm C}} 
\Big(\sum_{\langle i,j \rangle_{i_{\rm C}}} \sigma_i \sigma_j + 
 \sum_{\langle\langle i,k \rangle\rangle_{i_{\rm C}}, i \in A, k \in B}^{2L} 
\frac{1}{2} (\sigma_i y  + x  \sigma_k -xy) \notag  \\  
& +  
\sum_{\langle\langle i,k \rangle\rangle_{i_{\rm C}}, i \in B, k \in A}^{2L} \frac{1}{2} (\sigma_i x + y  \sigma_k -xy) \Big)  \notag \\ 
&=J \sum_{i_{\rm C}=1}^{N_{\rm C}} 
\Big(\sum_{\langle i,j \rangle_{i_{\rm C}}} \sigma_i \sigma_j + 
 \sum_{\langle\langle i,k \rangle\rangle_{i_{\rm C}}, i \in A}^{2L} 
  \sigma_i y  +   \sum_{\langle\langle i,k \rangle\rangle_{i_{\rm C}}, i \in B}^{2L}  \sigma_i x -2Lxy \Big), 
\label{H1CMF}
\end{align}
where $i \in$ A (B) means that $i$ belongs to sublattice A (B). 
Here $\sum_{\langle\langle i,k \rangle\rangle_{i_{\rm C}}, i \in A} \sigma_i y $, etc.\ means that the summation is taken over pairs at borders between site $i$ and neighboring site $k$ whose average $\langle \sigma_k \rangle$ is replaced by $y$, etc. 

Noting the relation $\langle \sigma_i \sigma_k \rangle=\langle \sigma_i\rangle \langle \sigma_k\rangle$ for sites $i$ and $k$ that belong to different clusters,  we obtain  
\begin{align}
\langle {\cal H}_1-{\cal H}_{1,\rm CMF}(x,y) \rangle &=J N_{\rm C}
\Big( \sum_{\langle\langle i,k \rangle\rangle_{i_{\rm C}}, i \in A}^{2L}  
\frac{1}{2} \big(\langle \sigma_i \rangle \langle \sigma_k \rangle- 2y\langle \sigma_i \rangle  \big) \\
&+ \sum_{\langle\langle i,k \rangle\rangle_{i_{\rm C}}, i \in B}^{2L} 
\frac{1}{2} \big(\langle \sigma_i \rangle \langle \sigma_k \rangle-2x \langle \sigma_i \rangle \big)  +2Lxy \Big). \notag
\label{DH1}
\end{align}
In the practical calculation, $\langle \sigma_k \rangle$ in a neighboring cluster is replaced by $\langle \sigma_{k'} \rangle$ at equivalent site $k'$ in the ${i_{\rm C}}$th cluster (see Fig.~\ref{Fig1_Cluster}), i.e., $\langle \sigma_{k'} \rangle=\langle \sigma_k \rangle$.

Next, we construct a CMF Hamiltonian for ${\cal H}_2$. 
${\cal H}_2$ is rewritten as 
\begin{align}
{\cal H}_2 = &-\frac{A}{2N} \Big(\sum_{i_{\rm C}}^{N_{\rm C}} \sum_i^{L^2} \sigma_i \Big)\Big(\sum_{j_{\rm C}}^{N_{\rm C}} \sum_j^{L^2} \sigma_j \Big) \\
& -H \sum_{i_{\rm C}}^{N_{\rm C}} \sum_i^{L^2} \sigma_i.  \notag
\label{H2}
\end{align}
By the following replacements, 
\begin{align}
\sum_i^{L^2}\sum_j^{L^2} \sigma_i \sigma_j=\sum_{i \in A}^{L^2/2}\sum_{j \in A}^{L^2/2}  \sigma_i \sigma_j+\sum_{i \in A}^{L^2/2}\sum_{j \in B}^{L^2/2}  \sigma_i \sigma_j+\sum_{i \in B}^{L^2/2}\sum_{j \in A}^{L^2/2}  \sigma_i \sigma_j+\sum_{i \in B}^{L^2/2}\sum_{j \in B}^{L^2/2}  \sigma_i \sigma_j, 
\end{align}
\begin{align}
\sum_{i \in A}^{L^2/2}\sum_{j \in A}^{L^2/2}  \sigma_i \sigma_j\simeq \sum_{i \in A}^{L^2/2}\sum_{j \in A}^{L^2/2}  (\sigma_i x +x\sigma_j -x^2),
\end{align}
\begin{align}
\sum_{i \in A}^{L^2/2}\sum_{j \in B}^{L^2/2}  \sigma_i \sigma_j \simeq \sum_{i \in A}^{L^2/2}\sum_{j \in B}^{L^2/2}  (\sigma_i y +x\sigma_j -xy),
\end{align}
\begin{align}
\sum_{i \in B}^{L^2/2}\sum_{j \in A}^{L^2/2}  \sigma_i \sigma_j \simeq \sum_{i \in B}^{L^2/2}\sum_{j \in A}^{L^2/2}  (\sigma_i x +y\sigma_j -xy),
\end{align}
and 
\begin{align}
\sum_{i \in B}^{L^2/2}\sum_{j \in B}^{L^2/2}  \sigma_i \sigma_j \simeq \sum_{i \in B}^{L^2/2}\sum_{j \in B}^{L^2/2}  (\sigma_i y +y\sigma_j -y^2), 
\end{align}
 we have 
\begin{align}
\sum_i^{L^2}\sum_j^{L^2} \sigma_i \sigma_j & \simeq \frac{L^2}{2} \sum_{i}^{L^2} \sigma_i(x+y)+\frac{L^2}{2} \sum_{j}^{L^2} (x+y)\sigma_j-(x+y)^2L^4/4. 
\end{align}

Then we obtain a CMF Hamiltonian ${\cal H}_{2,\rm CMF}(x,y)$ for ${\cal H}_2$ as 
\begin{align}
{\cal H}_{2,\rm CMF}(x,y) &=-\frac{A}{2N} \sum_{i_{\rm C}}^{N_{\rm C}} \sum_{j_{\rm C}}^{N_{\rm C}}\Big( \frac{L^2}{2} \sum_{i}^{L^2} \sigma_i(x+y)+\frac{L^2}{2} \sum_{j}^{L^2} (x+y)\sigma_j -(x+y)^2L^4/4 \Big) \notag \\
&-H \sum_{i_{\rm C}}^{N_{\rm C}} \sum_i^{L^2} \sigma_i \notag \\
&= \sum_{i_{\rm C}}^{N_{\rm C}} \Big(-\frac{A}{2}  \sum_{i}^{L^2} \sigma_i(x+y) 
+\frac{A}{8} L^2 (x+y)^2  -H \sum_i^{L^2} \sigma_i      \Big). 
\label{H2CMF}
\end{align}
It is noted that the following relation holds.  
\begin{align}
\langle {\cal H}_2-{\cal H}_{2,\rm CMF}(x,y) \rangle &=- N_{\rm C} \frac{A}{2 L^2} \Big(\sum_i^{L^2} \langle \sigma_i \rangle \Big)\Big(\sum_j^{L^2} \langle \sigma_j \rangle \Big)  \\
&+N_{\rm C} \frac{A}{2}  \Big((x+y) \sum_i^{L^2} \langle \sigma_i\rangle -\big(\frac{x+y}{2}\big)^2 L^2 \Big)  \notag 
\label{DH2}
\end{align}
Finally, the CMF Hamiltonian is given by 
\begin{align}
{\cal H}_{\rm CMF}(x,y)&=H_{1,\rm CMF}(x,y)+ H_{2,\rm CMF}(x,y) \notag \\ 
&={\cal \tilde{H}}_{\rm CMF}(x,y)
+ \sum_{i_{\rm C}=1}^{N_{\rm C}} \Big( -2J Lxy + \frac{A}{8} (x+y)^2L^2 \Big),
\end{align}
where
\begin{align}
{\cal \tilde{H}}_{\rm CMF}(x,y)& 
\equiv  \sum_{i_{\rm C}=1}^{N_{\rm C}} \Bigg[ J \Big(   \sum_{\langle i,j \rangle_{i_{\rm C}}} \sigma_i \sigma_j +  \sum_{\langle\langle i,k \rangle\rangle_{i_{\rm C}}, i \in A}^{2L}  \sigma_i y  +   \sum_{\langle\langle i,k \rangle\rangle_{i_{\rm C}}, i \in B}^{2L}  \sigma_i x \Big) \\
&- \frac{A}{2}(x+y) \sum_{i}^{L^2} \sigma_i -H  \sum_i^{L^2} \sigma_i  \Bigg]. \notag
\end{align}

Noting that $\langle {\cal H}-{\cal H}_{\rm CMF}(x,y) \rangle=\langle {\cal H}_1-{\cal H}_{1,\rm CMF}(x,y) \rangle+\langle {\cal H}_2-{\cal H}_{2,\rm CMF}(x,y) \rangle$, we have the variational free energy:  
\begin{align}
F_{\rm v}&=-\frac{1}{\beta} \ln \mathrm{Tr} \exp ({-\beta  {\cal H}_{\rm CMF} }) + \langle {\cal H} -{\cal H}_{\rm CMF} \rangle   \notag \\
&=-\frac{1}{\beta} \ln \mathrm{Tr} \exp ({-\beta  {\cal \tilde{H}}_{\rm CMF} }) + N_{\rm C}
\Big( \sum_{\langle\langle i,k \rangle\rangle_{i_{\rm C}}, i \in A}^{2L}  
\frac{1}{2} J \big(\langle \sigma_i \rangle \langle \sigma_k \rangle- 2y\langle \sigma_i \rangle  \big)\\
&+ \sum_{\langle\langle i,k \rangle\rangle_{i_{\rm C}}, i \in B}^{2L} 
\frac{1}{2} J \big(\langle \sigma_i \rangle \langle \sigma_k \rangle-2x \langle \sigma_i \rangle \big) -
\frac{A}{2L^2} \sum_i^{L^2} \langle \sigma_i \rangle \sum_j^{L^2} \langle \sigma_j \rangle +\frac{A}{2} (x+y) \sum_i^{L^2} \langle \sigma_i\rangle  \Big).  \notag
\end{align}

\subsection{Variational equations}

The variational equations for the free energy are given by  
\begin{align}
\partial_x F_{\rm v}(x,y)=0 \;\;\; {\rm and} \;\;\; \partial_y F_{\rm v}(x,y)=0 ,
\label{eq:vareq}
\end{align}
and it follows that  
\begin{align}
&-\frac{1}{\beta} \partial_x \ln \tilde{Z} + \frac{J}{2} \sum_{\langle\langle i,k \rangle\rangle_{i_{\rm C}}, i \in A}^{2L}  \Big( \big(\partial_x \langle \sigma_i \rangle \big) \langle \sigma_k \rangle + \langle \sigma_i \rangle \partial_x  \langle \sigma_k \rangle -2y \partial_x \langle \sigma_i \rangle \Big) \\
&+ \frac{J}{2} \sum_{\langle\langle i,k \rangle\rangle_{i_{\rm C}}, i \in B}^{2L}  \Big( \big( \partial_x \langle \sigma_i \rangle  \big) \langle \sigma_k \rangle + \langle \sigma_i \rangle \partial_x  \langle \sigma_k \rangle -2\langle \sigma_i \rangle -2x \partial_x \langle \sigma_i \rangle \Big) \notag \\
&- \frac{A}{L^2} \sum_i^{L^2} \langle \sigma_i \rangle \partial_x \sum_i^{L^2} \langle \sigma_i \rangle + \frac{A}{2}\sum_i^{L^2} \langle \sigma_i \rangle + A\frac{x+y}{2}\partial_x \sum_i^{L^2} \langle \sigma_i \rangle =0 \notag
\end{align}
and 
\begin{align}
&-\frac{1}{\beta} \partial_y \ln \tilde{Z} +\frac{J}{2} \sum_{\langle\langle i,k \rangle\rangle_{i_{\rm C}}, i \in A}^{2L}  \Big( \big( \partial_y \langle \sigma_i \rangle \big) \langle \sigma_k \rangle + \langle \sigma_i \rangle \partial_y  \langle \sigma_k \rangle -2 \langle \sigma_i \rangle -2y \partial_y \langle \sigma_i \rangle  \Big) \\
&+ \frac{J}{2} \sum_{\langle\langle i,k \rangle\rangle_{i_{\rm C}}, i \in B}^{2L}  \Big( \big( \partial_y \langle \sigma_i \rangle \big) \langle \sigma_k \rangle + \langle \sigma_i \rangle \partial_y  \langle \sigma_k \rangle -2x \partial_y \langle  \sigma_i \rangle \Big) \notag \\
&- \frac{A}{L^2} \sum_i^{L^2} \langle \sigma_i \rangle \partial_y \sum_i^{L^2} \langle \sigma_i \rangle + \frac{A}{2}\sum_i^{L^2} \langle \sigma_i \rangle + A\frac{x+y}{2}\partial_y \sum_i^{L^2} \langle \sigma_i \rangle =0,   \notag
\end{align}
where 
%\begin{align}
%\tilde{Z}^{{N}_{\rm C}}= \mathrm{Tr} \exp(-\beta {\cal \tilde{H}}_{\rm CMF}(x,y%)).  
%\end{align}
\begin{align}
\tilde{Z}= \{ \mathrm{Tr} \exp [-\beta {\cal \tilde{H}}_{\rm CMF}(x,y) ] \}^{1/{N}_{\rm C}}.  
\end{align}

The simultaneous solutions of these equations correspond to the stationary points of 
the variational free-energy landscape (minima, maxima, and saddle points). 

We solve these variational equations numerically by the Newton-Raphson method. 
In each iteration step for solving the equations, we calculate $\tilde{Z}^{{N}_{\rm C}}$, $\langle \sigma_i \rangle$, etc. with the use of $x$ and $y$ obtained in the previous step. The 
simultaneous solutions for $x$ and $y$ are obtained as converged values.  

It should be noted that in solving the equations 
a transfer-matrix method is adopted to perform the summation over the $2^{L^2}$ states for the Trace. We have to repeat the calculation many times, and in the following analysis, we study up to $L=8$, which can be done in a realistic computational time. 

\begin{figure}
\centerline{
\includegraphics[clip,width=12.0cm]{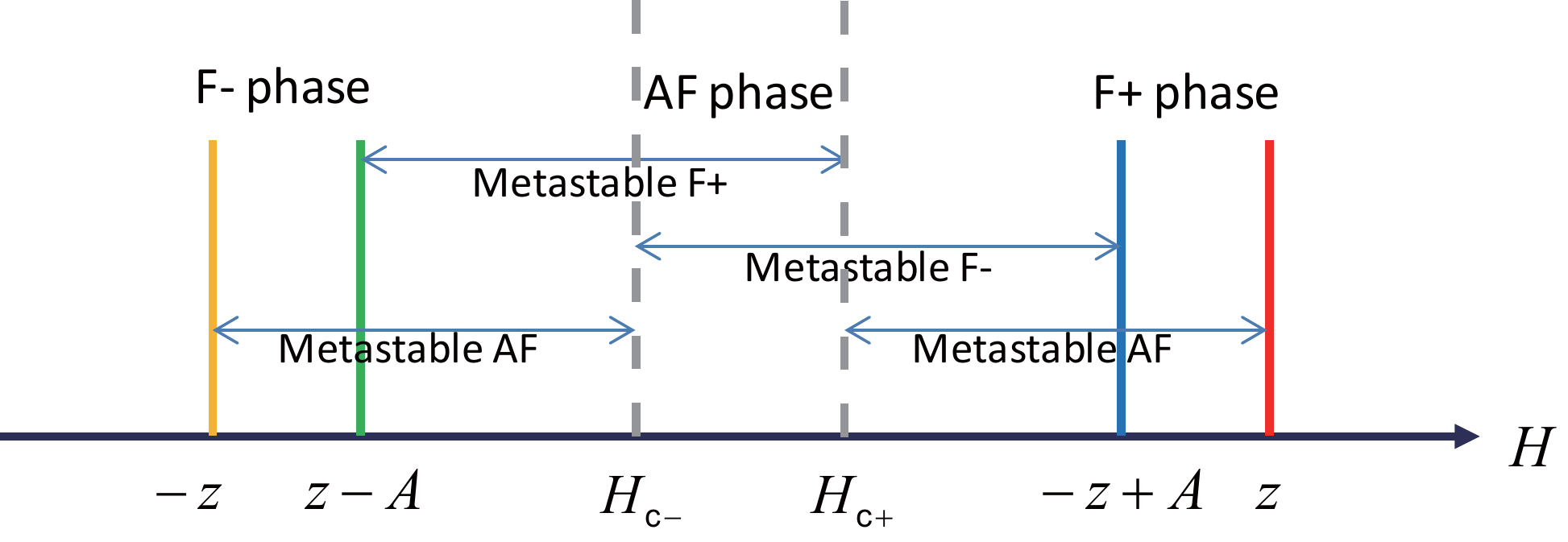}} 
\caption{Ground-state diagram for the model on a bipartite lattice with coordination number $z$ and $z<A<2z$. }
\label{Fig_GS}
\end{figure}

\section{Results and Discussion}
\label{sec_phase_diag}

Hereafter we take $J$ as the unit of energy and 
$H$, $A$, $T$, etc.\ are given in units of $J$. 
%The coordination number of the lattice is $z$. 

\subsection{Ground states}

In this subsection we discuss the ground-state diagram for the model on a bipartite lattice with 
coordination number $z$. In the rest of the paper we only consider a square lattice, for which 
$z=4$. 

At $T=0$ we can calculate stable phases explicitly in the limit $N \rightarrow \infty  $~\cite{Per1,Per3}. 
The per-site energy in the field $H$ of the fully ordered AF phase is 
given by $E_{\rm AFM}/N=-z/2$. 
That of the fully ordered F phase parallel to the field (F+) and that antiparallel to the field (F$-$) are given by $E_{+}/N=z/2-A/2-H$ and $E_{-}/N=z/2-A/2+H$, respectively. 
Thus, the transition field between the the AF and F+ phases is $H_{\rm c+}=z-A/2$, and that between the AF and F$-$ phases is  $H_{\rm c-} =-z+A/2$. 
If $A>2z$, the F phases are the ground states, while if $0 \le A<2z$ the AF phase is the ground state between $H_{\rm c+} =z-A/2$ and $H_{\rm c-}=-z+A/2$. 
Here we focus on the region $A<2z=8$ for $z=4$ as in our previous paper~\cite{Per1}.

Beside the fields at which the ground state changes (i.e., the first-order phase transition points), 
the limits of the metastable phases are also important. 
The limit of the metastability is estimated by calculating the field at which the excitation energy needed to nucleate a droplet of the equilibrium phase becomes zero. 
The excitation energy for a single flip from the F$-$ phase is $\Delta E=-2z-2H+2A$, and the upper 
limit of $H$ for the metastable F$-$ phase is $H=-z+A$. 
Thus the metastable region of the F$-$ phase exists in the region $H_{\rm c-}=-z+A/2 <H <-z+A$, and that of the F+ phase in the region $H_{\rm c+}=z-A/2 > H >z-A$ by symmetry. 
The excitation energy for a single down-spin flip from the AF phase is $\Delta E=2z-2H$, and $H=z$ is the upper limit for the metastable AF phase.
The metastable AF phase for $H>0$ exists in the region $H_{\rm c+}=z-A/2<H<z$ and 
that for $H<0$ in the region $-z<H<H_{\rm c-}=-z+A/2$ by symmetry. 
The ground-state diagram with $z<A<2z$ is depicted in Fig.~\ref{Fig_GS}. 
Hereafter we set $A=7.9$ and $z=4$. In this case $H_{\rm c+}=z-A/2=0.05$ and F$-$ and AF spinodal points are located at $H=-z+A=3.9$ and $H=z=4$, respectively.

\begin{figure}
\centerline{
\includegraphics[clip,width=8.0cm]{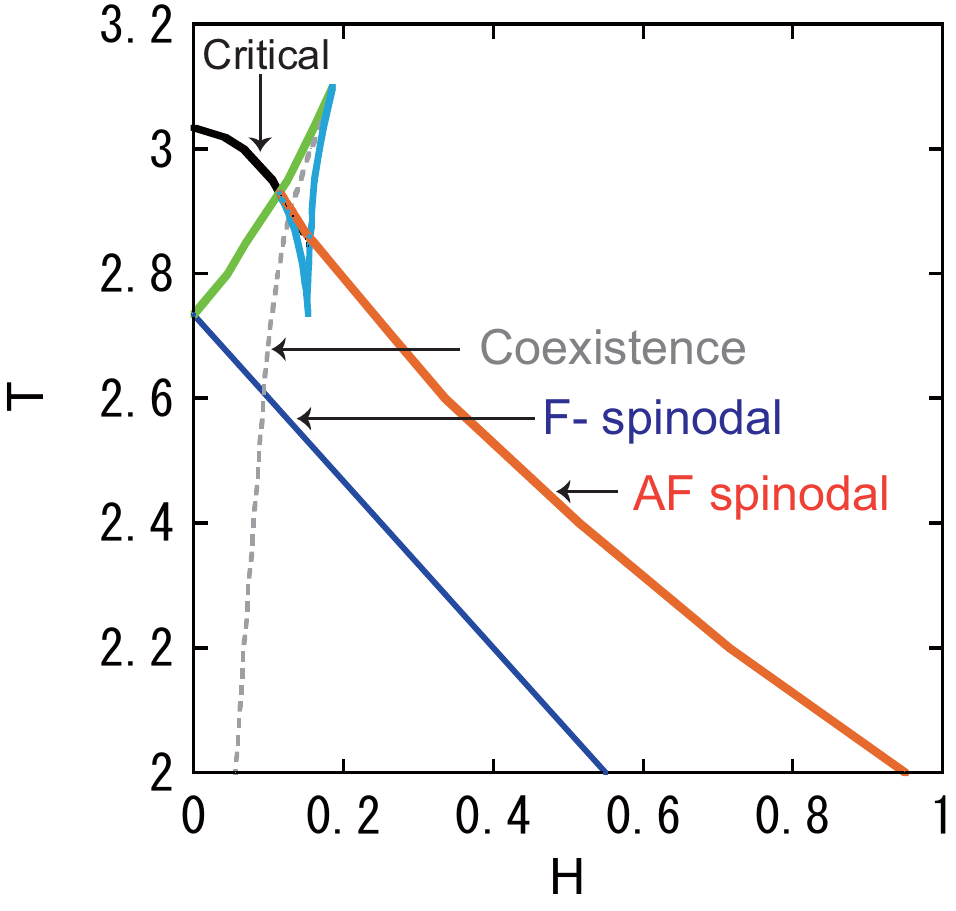}} 
\caption{Overview of the phase diagram of the CMF model with $z=4$ and $A=7.9$ for $L=6$.}
\label{Fig_overview}
\end{figure}

\subsection{Phase Diagram}

\begin{figure}
\centerline{
\includegraphics[clip,width=7.2cm]{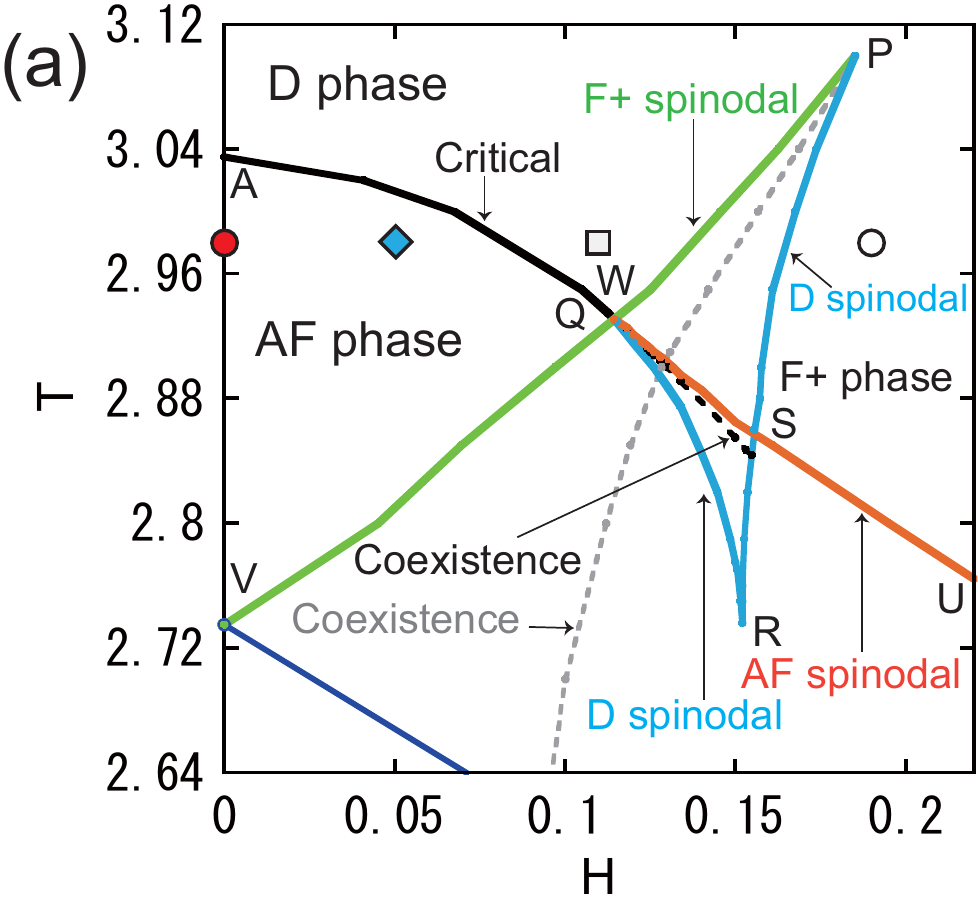}
\hspace{0.25cm} 
\includegraphics[clip,width=7.2cm]{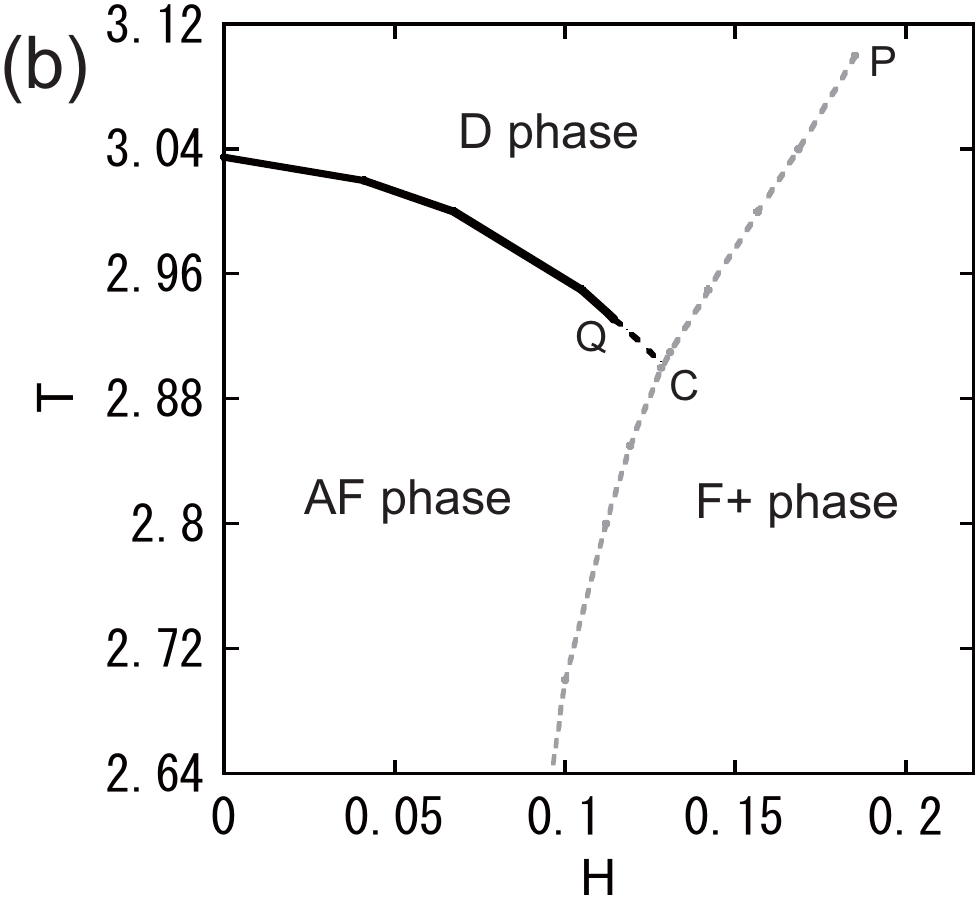} 
}
\vspace*{0.3cm} 
\centerline{
\hspace{0.25cm} 
\includegraphics[clip,width=7.5cm]{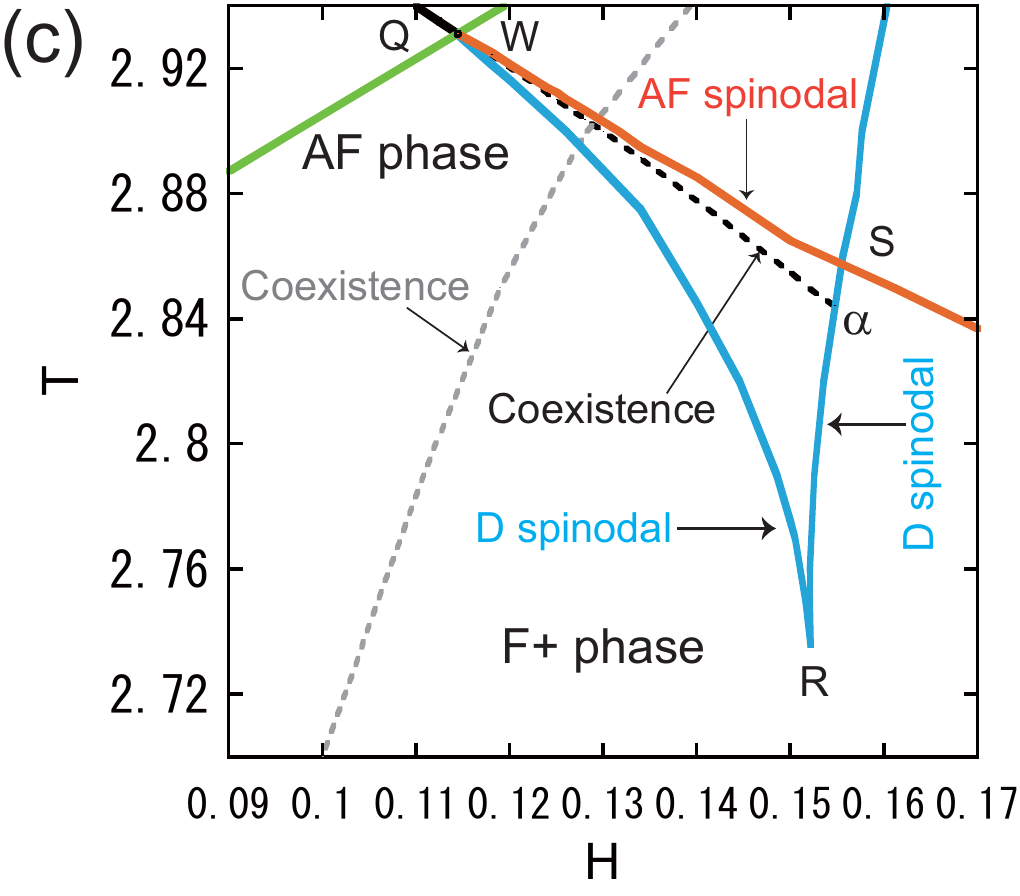} 
\hspace{0.0cm} 
\includegraphics[clip,width=7.5cm]{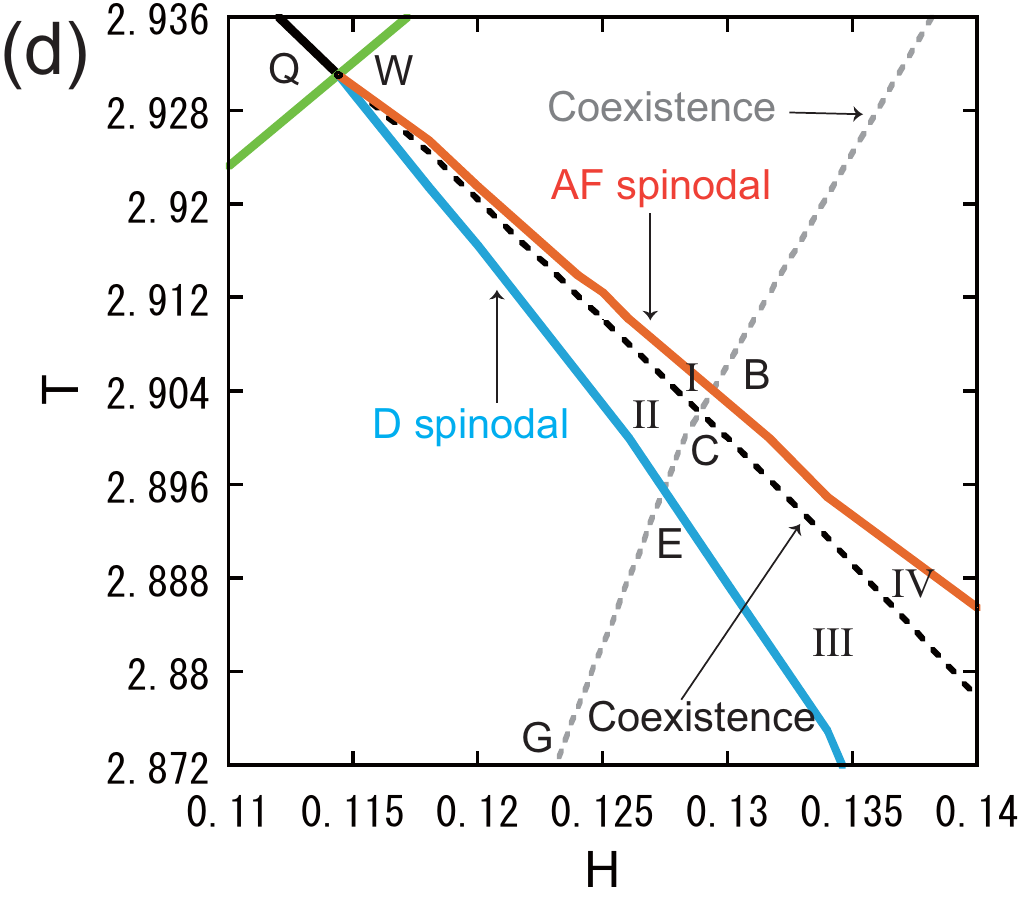} 
}
\caption{The phase diagram of the CMF model for $L=6$.
 (a) A horn region and its vicinity at relatively high temperatures. (b) The equilibrium phase diagram for (a). 
(c) Tri-stable regions and their vicinity. 
(d) Magnified detail of the tristable regions. See Fig.~\ref{Fig_regionV_VI} for magnified detail around point C. 
 }
\label{Fig2_Phase_diag_L6}
\end{figure}

The overview of the phase diagram at higher temperatures for $A=7.9$ and $L=6$ is shown in 
Fig.~\ref{Fig_overview}. This value of the long-range interaction strength $A$ is used for 
all the numerical results in this paper.  
The phase diagram is symmetric about the $T$ axis with an exchange between 
FM$+$ and FM$-$, and only the region $H \ge 0$ is shown. 
We note that the horn region in the phase diagram, which is not realized in the Bragg-Williams (BW) MF phase diagram (Fig.\ 1 in Ref.\ [1]), appears at relatively high temperatures. This indicates that the effect of the thermal fluctuations for systems with an interplay between competing SR and LR interactions can cause unusual structures in the phase diagram. 
Details of the horn structure in the phase diagram at higher temperatures  are shown for $L=6$ (Figs.~\ref{Fig2_Phase_diag_L6} (a)-(d)).

\begin{figure}
\centerline{
\includegraphics[clip,width=6.0cm]{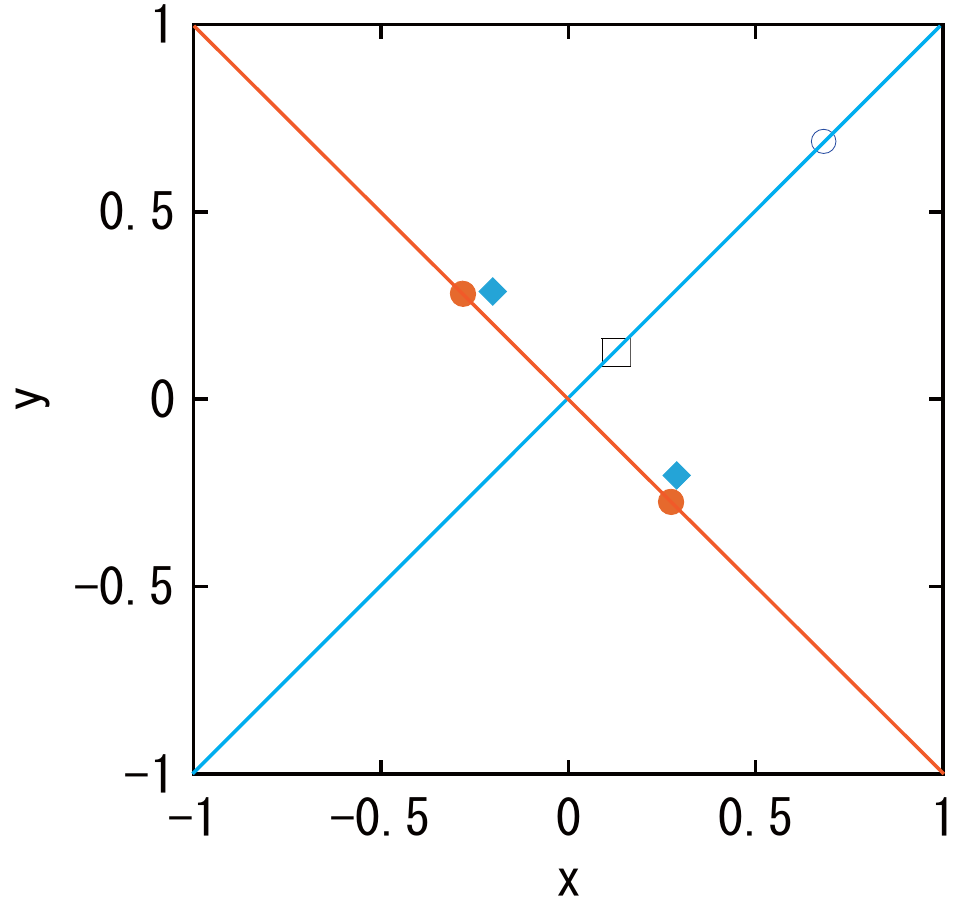}} 
\caption{Locations in $x$ and $y$ of global minima of $F_{\rm v}$ for different values of $H$ at T=2.98; 
 $(x,y)=(\pm 0.2781,\mp0.2781)$ at $H=0$ (red filled circles)  
and $(x,y)=(0.2884,-0.2023),(-0.2023,0.2884)$ at $H=0.05$ (blue filled diamonds), indicating the AF phase; 
$(x,y)=(0.1259,0.1259)$ at $H=0.11$ (open square) corresponding to the D phase;  and $(x,y)=(0.6847,0.6847)$ at $H=0.19$ (open circle), corresponding to the F+ phase.}
\label{Fig_ODspace}
\end{figure}

\begin{figure}
\centerline{
\includegraphics[clip,width=7.5cm]{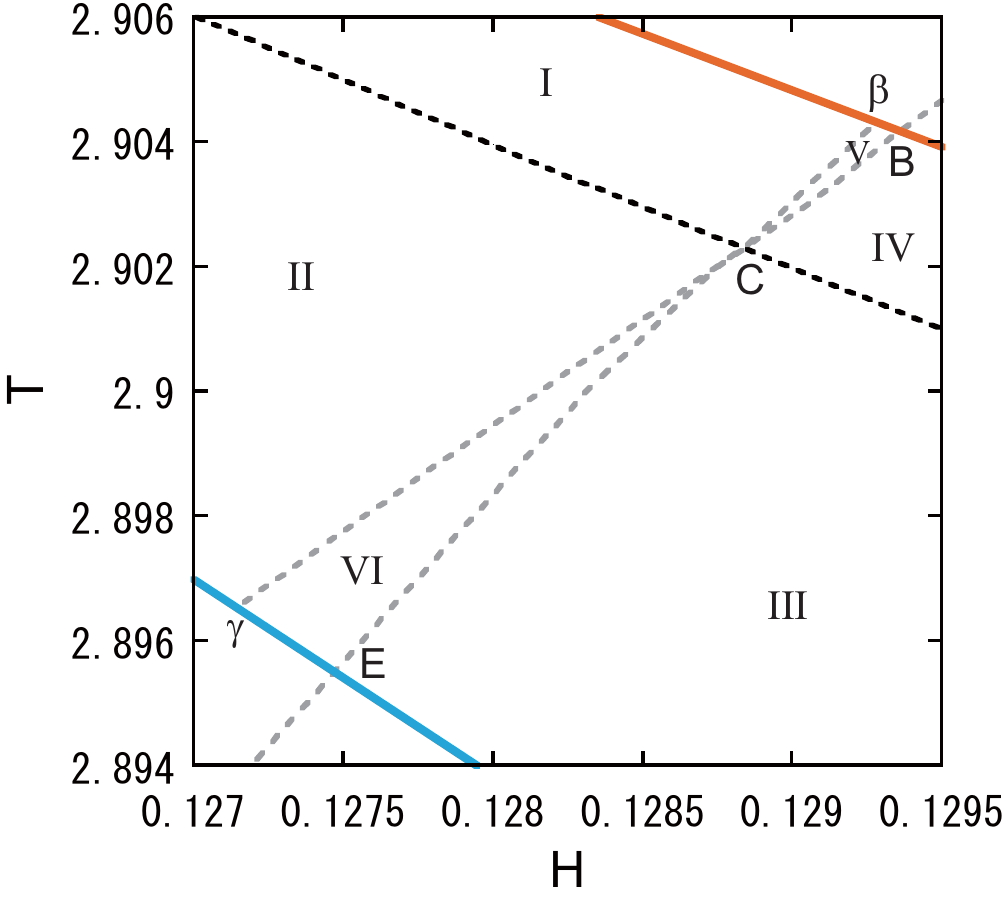}} 
\caption{ Magnified detail for regions V and VI in the tristable regions.}
\label{Fig_regionV_VI}
\end{figure}

\subsubsection{Variational parameter space}

Here we identify the phases in the space of the variational parameters, ($x$,$y$). In Fig.~\ref{Fig_ODspace} we schematically show the locations of the global free-energy minima at $T=2.98$, indicated by the symbols: filled circle, diamond, square, and open circle in Fig.~\ref{Fig2_Phase_diag_L6} (a). 

Decreasing the temperature at $H=0$ causes a separation of the D phase  ($(x,y)=(0,0)$) into the AF phases, $(x,y)=(\delta,-\delta)$ or $(-\delta,\delta)$. 
Eq.~\eqref{Ham2} does not contribute to the AF ordering at $H=0$, and 
 the critical temperature at $H=0$ corresponds to that of the pure Ising model. Namely,  $T_{\rm c}=z=4$ in the BW MF theory, while the exact value is $T_{\rm c} \simeq 2.269$~\cite{Onsager}. 
Here, the critical temperature $T_{\rm c}=3.035$ in the CMF model is closer to the exact 
$T_{\rm c}$ than the BW MF critical point. 
At $H=0$ and $T=2.98$, $(x,y)=(\pm 0.2781,\mp0.2781)$ are given by red filled circles in 
Figs.~\ref{Fig2_Phase_diag_L6} (a) and \ref{Fig_ODspace}.

If the field is increased up to $H=0.05$, these points move to the blue closed diamonds, located at $(x,y)=(0.2884,-0.2023),(-0.2023,0.2884)$. 
These minima representing the AF phase persist in the region below the line AQSU 
in the phase diagram.
Above the line, the stability of the AF phase vanishes, 
and the stable phases are uniformly magnetized phases, located on the line $y=x$ 
in Fig.~\ref{Fig_ODspace}. 
As to the characteristics of the horn structure, there are two distinguishable phases on the line of 
$y=x$: the D phase with a small $x(=y)$, e.g., 
$(x,y)=(0.1259,0.1259)$ at $H=0.11$, and the F+ phase with a large $x(=y)$, e.g., 
$(x,y)=(0.6847,0.6847)$ at $H=0.19$, which are given by the open square and circle in Fig.~\ref{Fig2_Phase_diag_L6}(a), respectively.

\subsubsection{Characteristics of the horn structure}

The D and F+ phases have the same symmetry but they are separated by a first-order phase transition. The point P is the critical point between F+ and D phases (Figs.~\ref{Fig2_Phase_diag_L6} (a) and (b)). 
The coexistence line between the F+ and D phases 
is located around the middle between the F+ and D spinodal lines, which is consistent with a very recent MC study~\cite{Per3} by the macroscopically constrained Wang-Landau method~\cite{Wang,Per2,Per4}. The location of the coexistence line was very close to that of the D spinodal line when the mixed start method was applied to identify it in a previous study by an 
importance-sampling MC method~\cite{Per1}, but the present observation supports that the coexistence line PC is located in the middle of the horn. 
The shape of the horn structure PWS is similar to that of the MC study for $A=7$ in 
Fig.~5(b) in Ref.\ \cite{Per3}. Here points W and S are the intersections between lines AU and PV and between AU and PR, respectively.  

We find differences in the phase diagram between the CMF and MC studies. 
In contrast to the MC studies~\cite{Per1,Per3}, tristable regions 
with one globally stable and two metastable phases are seen using 
the CMF method. We find a tricritical point Q and a triple point C (see Eq.~\eqref{triple}) in the CMF method. The points Q and W are closely located here but 
are independent. In principle, point Q can be a point on line AS (see Sec.~\ref{L-dep}). The line WS is identified as a critical line in the MC studies, but due to the multistability it is not in the CMF study. 
The green line PV is the limit of the metastable F+ phase, and the blue line PR is the limit of the metastable D phase. 
The line QR is also a limit of metastability (spinodal) of the D phase. The point R is a characteristic point of the border of the metastable D phase. 
Between lines QS and QR, a first-order phase transition line between the metastable D and AF phases exists, which is drawn by the black dotted line Q$\alpha$ (see Figs.~\ref{Fig2_Phase_diag_L6}(c) and (d)).   

On the dotted line Q$\alpha$ the free energy of the D phase ($F_{\rm v}({\rm D})$) and that of the AF phase ($F_{\rm v}({\rm AF})$) are the same, i.e., $F_{\rm v}({\rm D})=F_{\rm v}({\rm AF})$. 
It should be noted that the dotted line terminates at the point $\alpha$ where the local minimum of the free energy for the D phase disappears. 
This type of special point is one of the characteristics of the phase diagram with tristable regions. 

The coexistence line between the F+ and D phases is given by a line on which $F_{\rm v}({\rm F+})=F_{\rm v}({\rm D})$, which connect points P, C, and $\gamma$ (Fig.~\ref{Fig_regionV_VI}). 
In the same way the coexistence line between the F+ and AF phases is given by a line on which $F_{\rm v}({\rm AF})=F_{\rm v}({\rm F+})$, which is given by the line GC$\beta$  (Fig.~\ref{Fig2_Phase_diag_L6}(d) and Fig.~\ref{Fig_regionV_VI}). 
At points $\beta$ and $\gamma$, the local minima for the AF and D phases disappear, respectively, and these points are also special points. 

%At the cross point C between the lines P$\gamma$ and G$\beta$, the free energies of all the three %phases are the same, i.e.,

Numerically we find that the three lines, Q$\alpha$, P$\gamma$, and G$\beta$, cross at a single point, C, in agreement with Gibbs' Phase Rule. At this crossing point, the free energies of all the three phases are the same, i.e.,
\beq
F_{\rm v}( {\rm F+})= F_{\rm v}({\rm AF})= F_{\rm v}({\rm D}) \quad {\rm at \;\;\; point \;\; C} \;.
\label{triple}
\eeq
The point C is {\it the triple point} of the phase diagram.

\subsubsection{Tristable regions in the phase diagram}

Now we characterize the tristable regions QRS, in which six regions exist, characterizing the 
relative stability of the three phases. 

\begin{description}
\item [Region I]: $F_{\rm v}($\rm D$) < F_{\rm v}($\rm AF$) < F_{\rm v}($\rm F+$)$. \\
The D phase is stable, the AF phase is metastable, and the F+ phase is secondary metastable. It is surrounded by line Q$\beta$C. 
\item [Region II]: $F_{\rm v}($\rm AF$) < F_{\rm v}($\rm D$) < F_{\rm v}($\rm F+$)$. \\
The AF phase is stable, the D phase is metastable, and the F+ phase is secondary metastable. It is surrounded by line Q$\gamma$C. 

\item [Region III]: $F_{\rm v}($\rm F+$) < F_{\rm v}($\rm AF$) < F_{\rm v}($\rm D$)$. \\ The F+ phase is stable, the D phase is metastable, and the AF phase is secondary metastable. It is surrounded by line REC$\alpha$. 

\item [Region IV]: $F_{\rm v}($\rm F+$) < F_{\rm v}($\rm D$) < F_{\rm v}($\rm AF$)$. \\
The F+ phase is stable, the AF phase is metastable, and the D phase is secondary metastable. It is surrounded by line SBC$\alpha$. 

\item [Region V]:
$F_{\rm v}($\rm D$) < F_{\rm v}($\rm F+$) < F_{\rm v}($\rm AF$)$. \\
The D phase is stable, the F phase is metastable, and the AF phase is secondary metastable. It is surrounded by line $\beta$BC (small region). 

\item [Region VI]:
$F_{\rm v}($\rm AF$) < F_{\rm v}($\rm F+$) < F_{\rm v}($\rm D$)$. \\
The AF phase is stable, the F+ phase is metastable, and the D phase is secondary metastable. It is surrounded by line $\gamma$EC (small region). 
 
\end{description}

\begin{figure}
\centerline{
\includegraphics[clip,width=17cm]{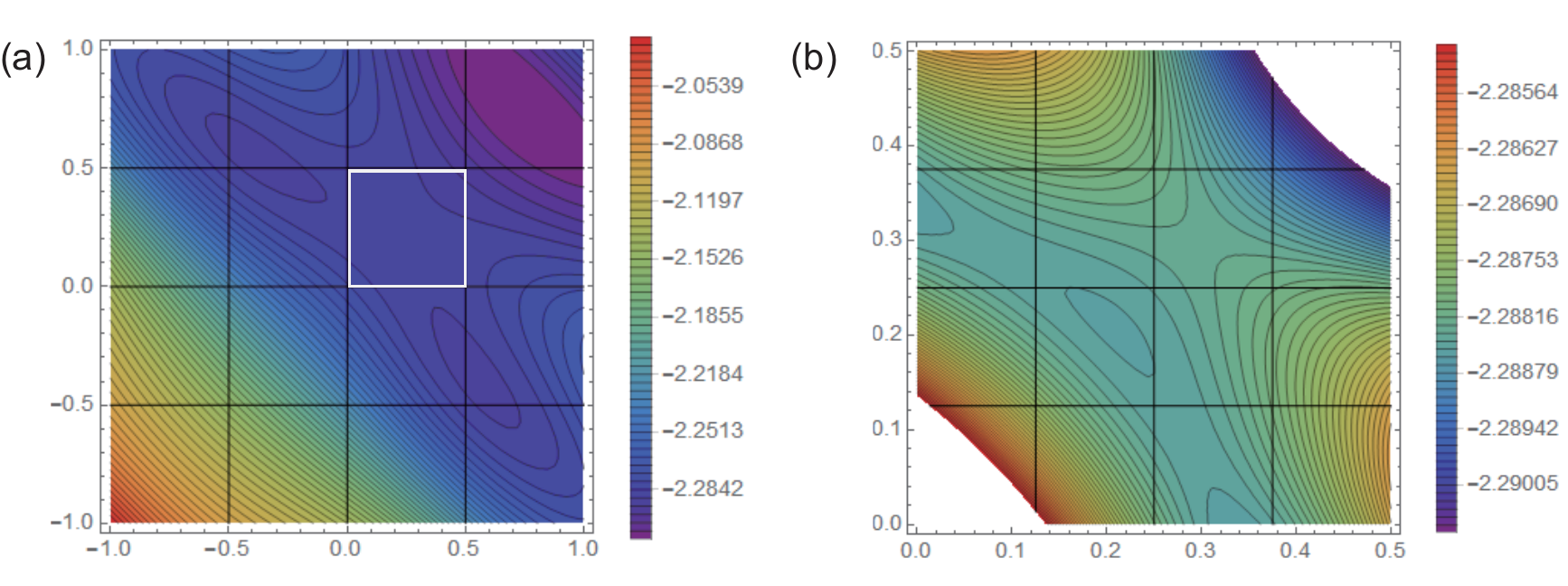}
}
\caption{(a) Contour plot of the per-site free energy in the $x$-$y$ plane at $(T,H)=(2.820,0.1490)$ in region III in the cluster MF model for $L=6$. 
%The thick, curved lines are the solutions of the variational equations, 
%$\partial F_{\rm v}(x,y) / \partial (x+y)$ (black) and $\partial F_{\rm v}(x,y) / \partial (x-y)$ (gray). 
(b) Magnified detail of the plot around the secondary metastable D phase in the region of the white square in (a). 
 }
\label{Fig3_contour_regionIII}
\end{figure}

\begin{figure}
\centerline{
\includegraphics[clip,width=17cm]{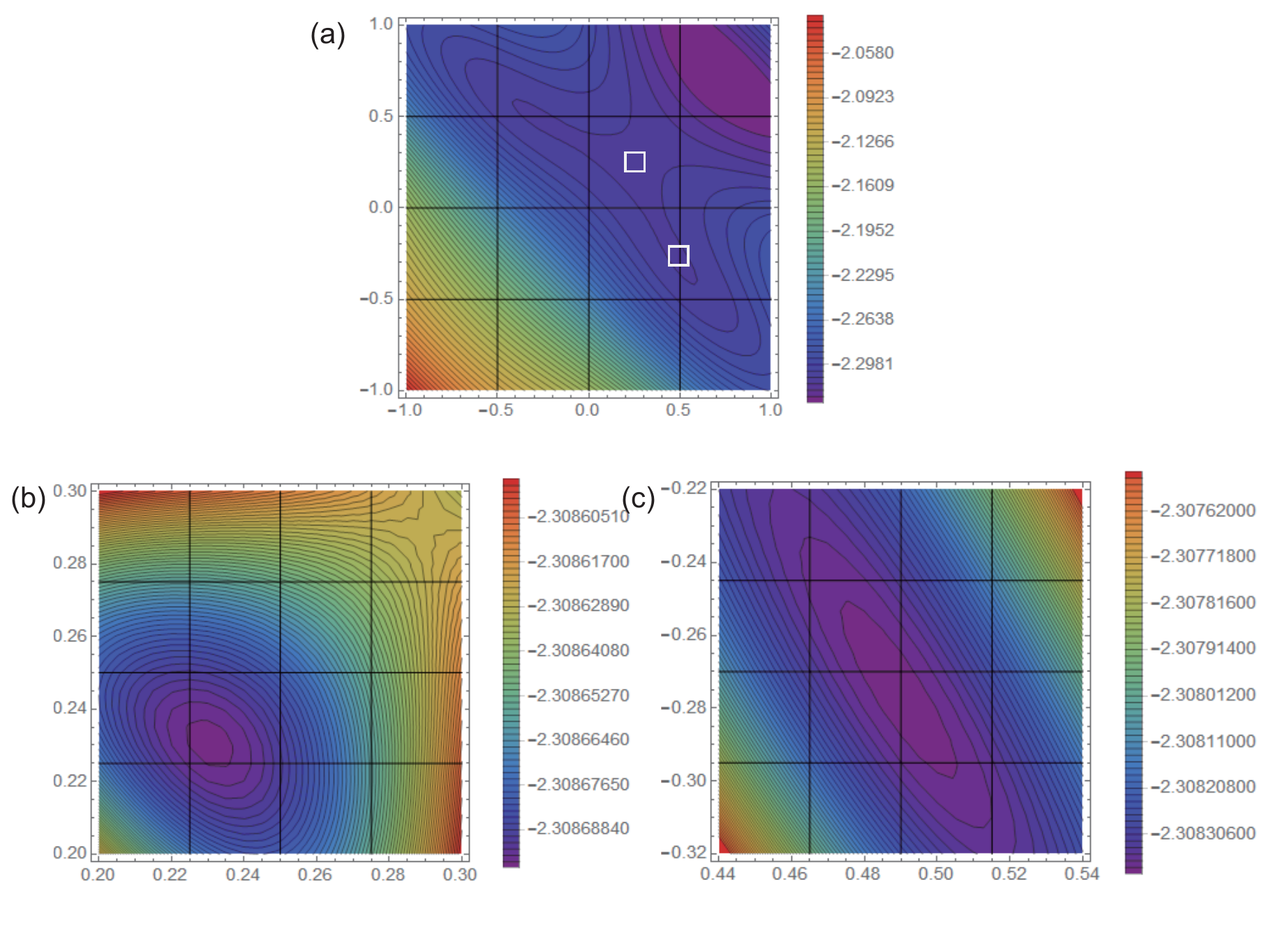}
}
\caption{(a) Contour plot of the per-site free energy in the $x$-$y$ plane at $(T,H)=(2.854,0.1530)$ in region IV in the cluster MF model for $L=6$. 
(b) Magnified detail of the plot around the metastable D phase in the region of the upper white square in (a). 
(c) Magnified detail of the plot around the secondary metastable AF phase in the region of the lower white square in (a). 
}
\label{Fig4_contour_regionIV}
\end{figure}

\subsection{Free-energy contour plots for tristable regions}

The analysis of contour plots of the free energy is helpful to understand the multistability of the model. 
For tristable phases three types of local minima, i.e., AF, F, and D phases, exist. 
To see this situation, we study tristability in regions III and IV. 
In Fig.~\ref{Fig3_contour_regionIII} we show contour plots of the per-site free energy, $F_{\rm v}/N$, in the $x$-$y$ plane at $(T,H)=(2.82,0.149)$ in region III.
On a rough scale (Fig.~\ref{Fig3_contour_regionIII}(a)), we can see minima only for AF and F+. 
The stable F+ phase is located at $(x,y)=(0.7550,0.7550)$, and the metastable AF phase is located at $(x,y)=(0.5623,-0.4022)$ (and $(x,y)=(-0.5623,0.4022)$) in Fig.~\ref{Fig3_contour_regionIII}(a).
However, in the magnified diagram (Fig.~\ref{Fig3_contour_regionIII}(b)), 
we find another local minimum located at $(x,y)=(0.2078,0.2078)$, which 
corresponds to the secondary metastable D phase. 

Although three local minima exist, we see that 
the energy barrier between the D and AF phases is much smaller than 
that between the F and AF phases -- at most on the order of the separation between the 
contours in Fig.~\ref{Fig3_contour_regionIII}(b), i.e., $\Delta F_{\rm v}/N \sim 9 \times 10^{-5}$. 
Therefore, we expect that the metastability of the D phase will be very difficult to detect by MC methods.  
An alternative rendition of the free-energy landscape shown in Fig.~\ref{Fig3_contour_regionIII} 
in terms of the sublattice magnetizations is discussed in the Appendix. 

We also depict the contour plot at $(T,H)=(2.854,0.1530)$ in region IV in  Fig.~\ref{Fig4_contour_regionIV}. 
Here the stable phase is F+ as well, located at $(x,y)=(0.7383,0.7383)$ (Fig.~\ref{Fig4_contour_regionIV}(a)), but the metastable D phase, located at $(x,y)=(0.2306,0.2306)$ in a magnified plot in Fig.~\ref{Fig4_contour_regionIV}(b), and the secondary metastable AF phase, located at $(x,y)=(0.4899, -0.2762)$ (and $(-0.4899, 0.2762)$) in a magnified plot in Fig.~\ref{Fig4_contour_regionIV}(c), are also realized. 
%There are also three minima. 
Here the energy barrier between the D and AF phases is much smaller than 
that between the F and AF phases.

\subsection{$L$ dependence}
\label{L-dep}

\begin{figure}
\centerline{
\includegraphics[clip,width=7.2cm]{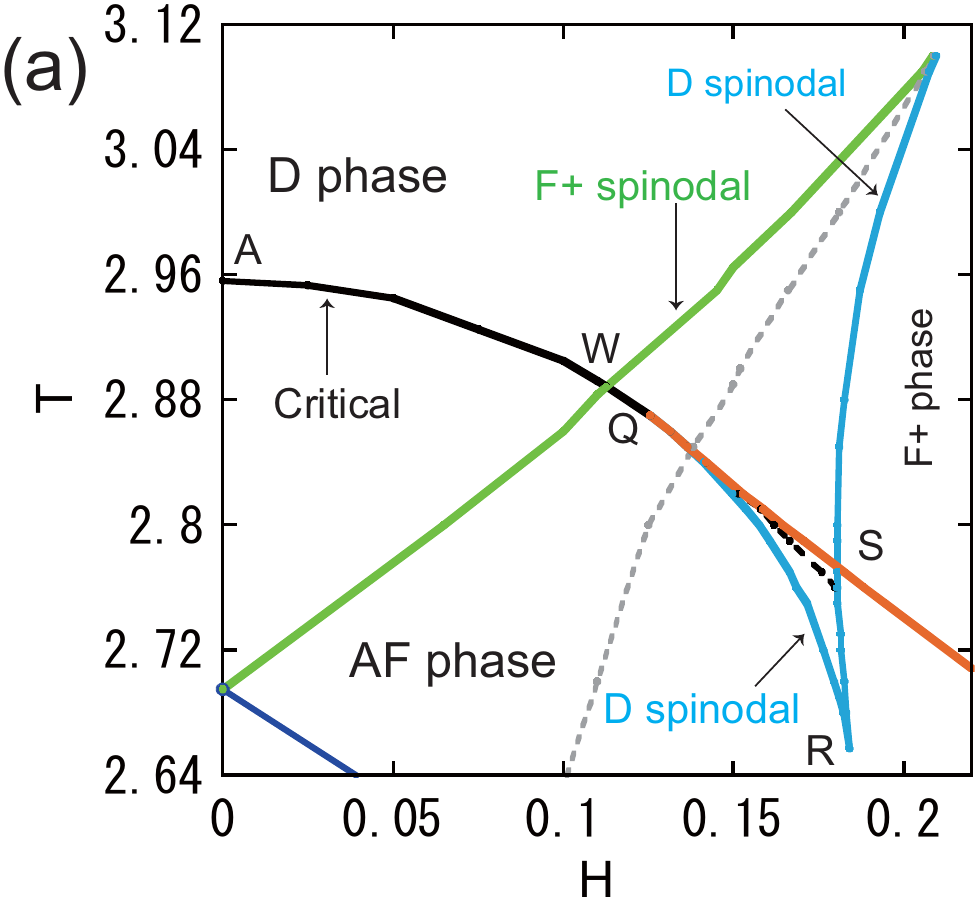}
\hspace{0.5cm} 
\includegraphics[clip,width=7.2cm]{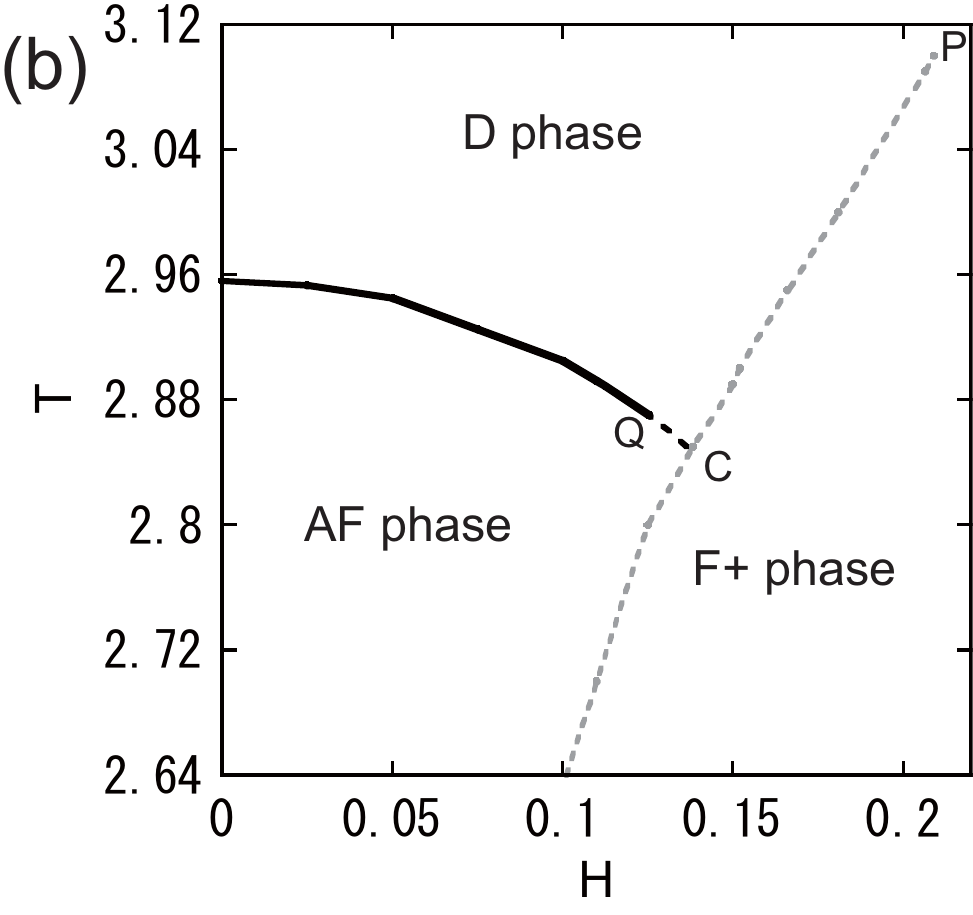}
}
\vspace*{0.3cm} 
\centerline{
\hspace{0.2cm} 
\includegraphics[clip,width=7.5cm]{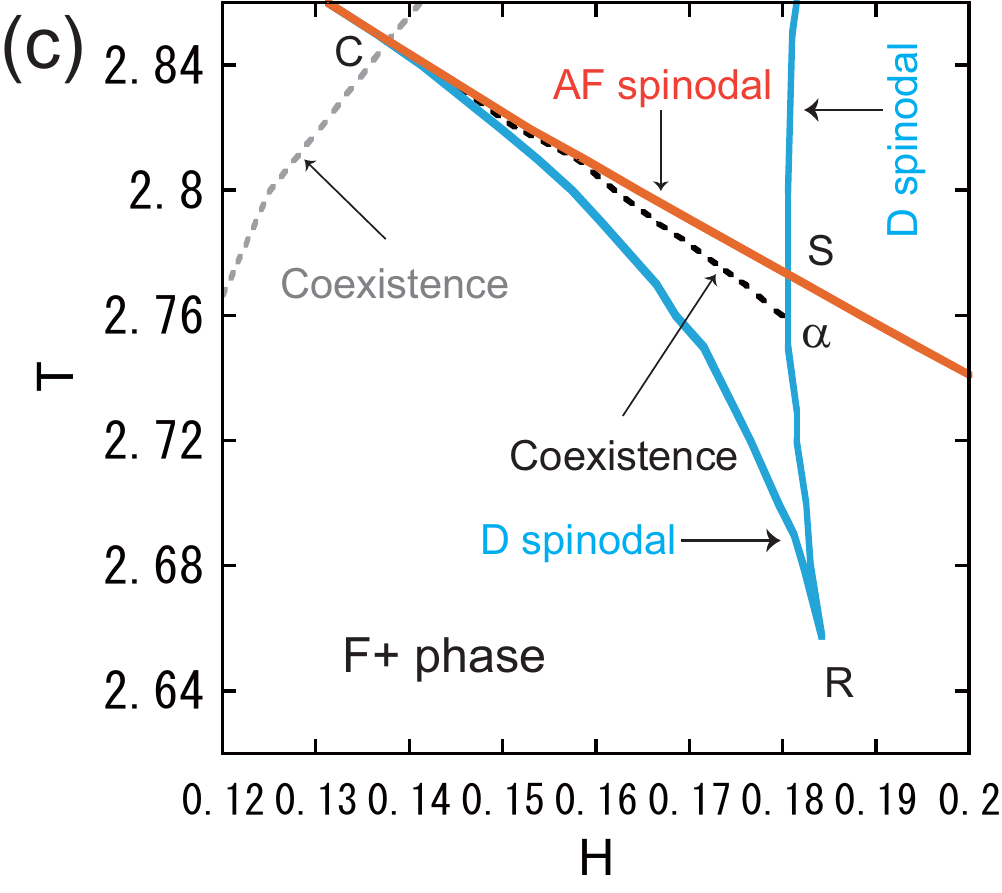}
\hspace{0.3cm} 
\includegraphics[clip,width=7.5cm]{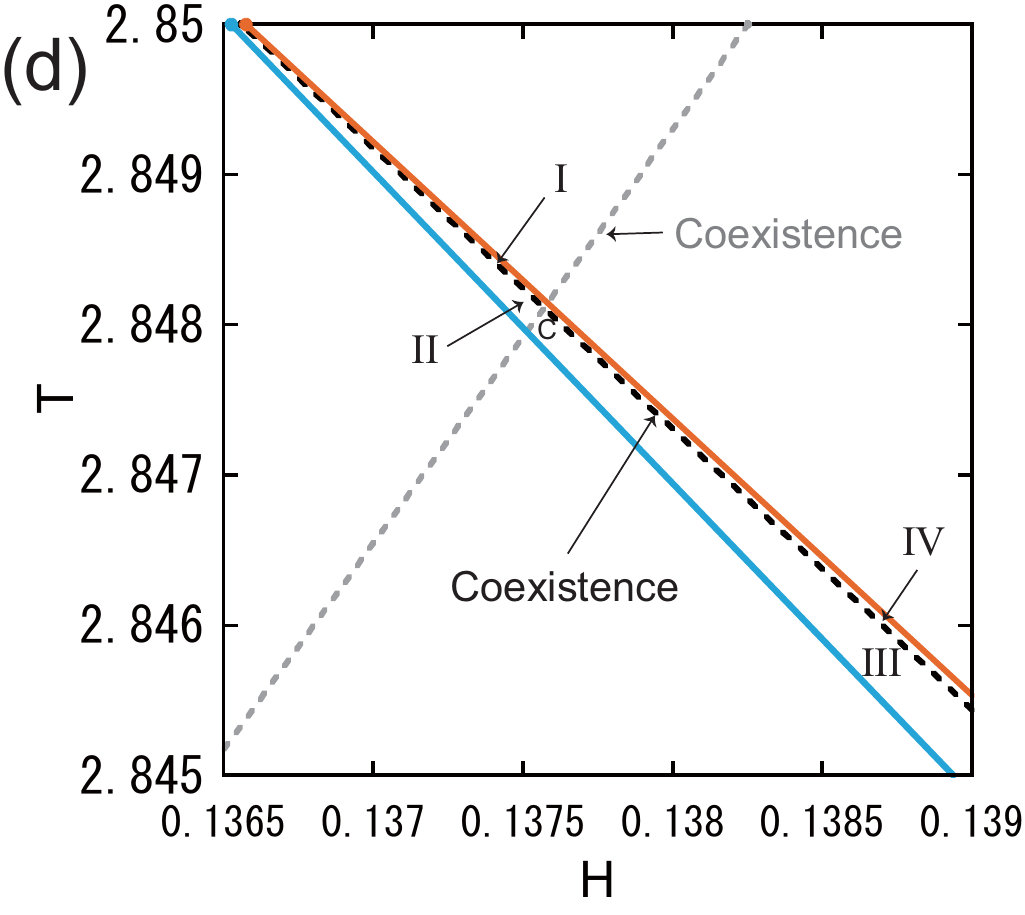}
}
\caption{The phase diagram of the CMF model for $L=8$. (a) A horn region at relatively high temperatures and its vicinity. (b) The equilibrium phase diagram for (a). 
(c) Tri-stable regions and the vicinity. 
(d) Magnified detail of the tristable regions. 
Comparing with Fig.~\protect\ref{Fig2_Phase_diag_L6}, we note that the horn region for $L=8$ is significantly larger than for $L=6$. }
\label{Fig5_Phase_diag_L8}
\end{figure}

To see the dependence on $L$, we study the phase diagram for $L=8$, given in Fig.~\ref{Fig5_Phase_diag_L8}. 
The horn region is depicted in Fig.\ \ref{Fig5_Phase_diag_L8}(a). 
Fig.\ \ref{Fig5_Phase_diag_L8}(b) is the phase diagram only for the equilibrium phases.  Figs.\ \ref{Fig5_Phase_diag_L8}(c) and (d) are magnified plots for the tristable regions. Tristable regions are also found here (regions V and VI are very narrow and not shown), and the diagram is qualitatively the same as that for $L=6$. 
However, some quantitative changes are found. 
The critical point ($T \simeq 2.956$) for $L=8$ is closer to the exact value $T \simeq 2.269$ than that for $L=6$. For larger $L$, it should approach the exact value. Comparing Fig.~\ref{Fig2_Phase_diag_L6}(a) and Fig.~\ref{Fig5_Phase_diag_L8}(a) on the same scale, the horn region for $L=8$ 
is significantly larger than for $L=6$. 
This change is understood as follows. 

The thermal fluctuations more strongly affect the SR interaction for larger $L$, which 
leads to stabilization of the D phase rather than the AF phase. 
Thus the location of the phase boundary between the D and AF phases will shift to the low-temperature side.  
On the other hand, the LR interaction is less affected by the thermal fluctuations and it approaches the MF interaction for larger $L$. 

The main concern is the existence of the tristable regions in the limit $L\rightarrow\infty$. 
We compare Fig.\ \ref{Fig5_Phase_diag_L8}(c) and Fig.\ \ref{Fig2_Phase_diag_L6}(c) on the same scale, and find that regions I and II become smaller for the larger $L$, while regions III and IV still exist in a wide range. 
For larger systems sizes ($L$), the first-order transition (line Q$\alpha$) between the AF and D phases may be reduced. 
However, there is a possibility that the metastabilities of the AF and D phases, especially in regions III and IV exist for larger $L$. 

From the above-mentioned considerations, we may give three possible scenarios for the horn structure and its vicinity in the phase diagram.
\begin{description}
\item [Scenario 1:] the transition between the D and AF phases is of second order and that 
between the metastable D and metastable AF phases is also of second order. 
\item [Scenario 2:] 
the transition between the D and AF phases for line QC is of (weak) first order, 
and that between the metastable D and metastable AF phases (line C$\alpha$) is also of first order.
\item [Scenario 3:] 
the transition between the D and AF phases is of second order, in which there is no triple point (points B, C, and E become one point), and the transition  between the metastable D and metastable AF phases is of second order on line CQ and of first order on line Q$\alpha$. 
\end{description}

Scenario 1 is the conclusion in the MC studies~\cite{Per1,Per3}, while 
Scenario 2 corresponds to the result of the present study. 
The tricritical point Q is located closely to point W in the present study, but in principle point Q can be a point on line AS, and Scenario 3 is a possible speculation from the present study and the MC studies. 
If point Q coincides with point C, the transition between the D and AF phases is of second order and that between the metastable D and metastable AF phases is of first order. 
If point Q coincides with point $\alpha$, points Q, R, and S become one point and tristable regions disappear. Line QS becomes a critical line and 
it corresponds to Scenario 1. 

It will be practically impossible to distinguish between Scenarios 1, 2, and 3 by MC studies for 
finite systems. 
However, the cause of larger error bars in the Binder plot for the transition between the metastable D and metastable AF phases in the MC studies~\cite{Per3}, especially at higher fields, might 
possibly be attributed to scenario 2 or 3. 

Figures 5 and 14 in the MC study~\cite{Per3} show that between the parameter $A=7$ and $A=8$ the relation of the location between the critical line and the F$-$ (and F+) spinodal lines changes at lower fields, although the equilibrium phase diagram is qualitatively the same except at $H=0$. There, point A in Fig.~\ref{Fig2_Phase_diag_L6}(a) is located lower than point V. There is a critical value of $A$ between 7 and 8, defined as $A_{\rm c}$. 
Here we have studied the case of $A=7.9$ because the horn structure appears more easily than for $A=7$ and can be studied by the CMF method. 
Thus, rigorously the above scenarios can be applied for $A<A_{\rm c}$ in the 
thermodynamic limit. 

\section{Discussion and Summary}
\label{summary}

We studied the phase diagram of the IA model with infinitely long-range F interactions with the cluster mean-field method.
Evaluating the variational free energy and solving the variational equations of the free energy with the use of the transfer-matrix method, we presented detailed structures of the phase diagram, including tristable regions for AF, F and D phases. 

In contrast to the simple BW mean-field (MF) theory for two sublattices, applied to this model in 
Ref.~\cite{Per1}, we observed unusual horn structures similar to those 
obtained in MC studies~\cite{Per1,Per3,Nishino_AF2}. We found that the interplay between the thermal fluctuations of the SR interaction enabled by the multisite clusters 
and the mean-field nature caused by the LR interaction is an essential requirement for the realization of such unusual structures. 

 For larger $L$ in the CMF method, thermal fluctuation effects are included up to a range comparable to the system size. 
The fluctuation effects from the SR interactions are important 
for the stabilization of the D phase versus the AF phase, 
resulting in the shift of the critical line and line QC to the lower temperature side. 
On the other hand, the LR interaction is more robust against the fluctuations, especially for larger 
$L$, and the spinodal lines are stabilized by the LR interaction. These competing effects cause the 
tricritical points observed in the BW MF  phase diagram~\cite{Per1} to decompose into the 
unusual horn structures seen by both the CMF and MC methods. 

%{\color{blue} The BW MF phase diagram~\cite{Per1} is characterized by tricritical points, at which the critical line and F+ and AF spinodal lines merge.  
%but they cannot exit due to the reduction of the AF critical line to the low temperature side and instead the MF critical points appear as the tops of the horns accompanying new spinodal lines, i.e., the D spinodal lines.
%}  

At $H=0$, the total magnetization is zero and the system is equivalent to the 
Ising antiferromagnet which has the critical temperature $T\simeq2.27$ on the square lattice. This value should be realized in the large $L$ limit. Indeed, we found that for larger $L$ the phase boundary between the AF and D phases shifts to lower temperatures in the direction of the exact value. 
On the other hand, the phase boundary of the F+ and D phases shifts to the higher field side, and the temperature of the critical point (the merging point of the phase boundaries of the metastable F+ and D phases) i.e., the top of the horn structure, changes relatively little. Therefore, the horn region expands with increasing $L$. 
In the MC studies~\cite{Per1,Per3}, the MF critical points are located at $(H,T)\simeq(0.56,2.61)$ for $A=7$, while the critical temperature at $H=0$ is significantly lower at $T\simeq2.27$, resulting in prominent horn structures. We expect the locations of the horn regions in the CMF phase diagram to approach those of the MC phase diagram for larger $L$. 

We investigated in detail the phase boundaries around the horn regions and found six tristable regions located closely to the horn regions. 
The MC studies~\cite{Nishino_AF2,Per1,Per3} presented four different bistable regions around the horn regions, i.e., the D phase with the metastable F+ phase, the F+ phase with the metastable D phase, the AF phase with the metastable F+ phase, and the F+ phase with the AF metastable phase
. 
The six tristable regions seen in the present CMF study were not identified, and  the boundary between the AF and D phases (line AS) was classified as a critical line.
On the other hand, a tricritical point Q and a triple point C were found in the 
CMF study, as well as three characteristic points, $\alpha$, $\beta$, and $\gamma$, at which the D phase, the AF phase, and the D phase 
become unstable, respectively. 
This tristability might be the cause of the larger error bars in the Binder plots in the higher-field region in the MC study~\cite{Per3}. 

As the size dependence of the phase diagram suggested, if the tricritical point Q approaches point $\alpha$, points B, C, and E become the same point, and the line AQC is the critical line. 
This point (C) is the endpoint of the first-order transition between the AF and F+ phases and also that between the D and F+ phases, which is a characteristic point in the phase diagram with three order parameters. 

Here, we propose three possible scenarios for the phase transitions associated with the horn structure.  
However, the metastability between the AF and D phases is weak and it would be difficult to detect the tristable regions by finite-size MC studies. 

The phase diagram with the triple point could be regarded as a lattice model of three phases of materials, e.g., the gas, liquid and solid. So far, for the gas-liquid phase transition the ferromagnetic Ising model is used while for the liquid-solid phase transition an AF Ising model is used. However, lattice models for the full three phases are not known to our knowledge. The present model suggests that a lattice-gas model with short-range AF interactions and rather long-range F interactions may have a phase diagram including a triple point.

\section*{Acknowledgments}
The present work was supported by Grants-in-Aid for
Scientific Research C (No. 17K05508) 
from MEXT of Japan, and the Elements Strategy Initiative
Center for Magnetic Materials (ESICMM) under the outsourcing
project of MEXT.
The authors also thank the Supercomputer Center, the Institute for Solid State Physics, the University of Tokyo, for the use of the facilities. 
PAR is grateful for hospitality in the Department of Physics, Graduate School of Science, at the University of Tokyo. Work at Florida State University was supported in part by U.S.\ National Science Foundation Grant No.\ DMR-1104829. 
This work was partly supported by the Research Council of Norway through its Centres of Excellence funding scheme, Project No. 262644 (PoreLab).

\appendix*
\section{Free energy vs sublattice magnetizations}
\label{sec:FE}

\begin{figure}
\vspace*{-0.5cm}
\begin{center}
\includegraphics[angle=0,width=.5\textwidth]{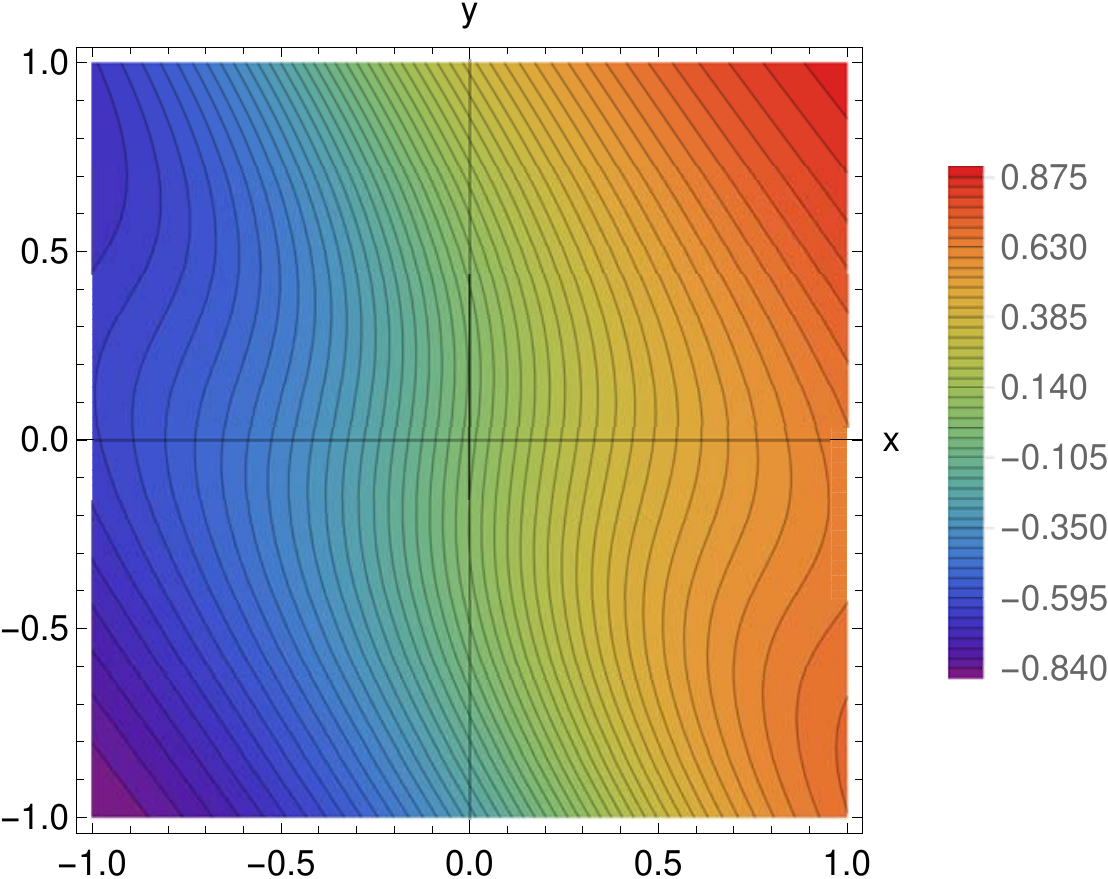}
\end{center}
\vspace*{-0.5cm}
\caption{
%\baselineskip=0.15truecm
Contour plot showing the dependence of the A sublattice magnetization $m_{\rm A}$ on the 
variational parameters, $x$ and $y$. 
The parameters are $A=7.9$ with $(T,H)=(2.820,0.149)$ and $L=6$.
The dependence on $x$ is monotonic but nonlinear, while 
the dependence on $y$ is nonmonotonic. The nonmonotonicity in the $y$-direction 
is caused by the LR interactions. It is not present for the pure Ising antiferromagnet, in which 
$A=0$. 
%The analogous plot for $m_{\rm B}$ is obtained from this one by reflection about the $y=x$ axis.  
 }
\label{fig:mAvsxy}
\end{figure}

In the variational MF approximation used here, stationary points of the free energy are obtained 
as the simultaneous solutions of the variational equations, Eq.~(\ref{eq:vareq}). The variables $x$ 
and $y$ are thus not order parameters of the system, but rather variational parameters. 
An example of the nonlinear and nonmonotonic dependence on $x$ and $y$ of 
the magnetization on the A sublattice, here defined as 
\begin{equation}
m_{\rm A} = \frac{2}{L^2} \sum_{i \in {\rm A}}^{L^2/2} \langle \sigma_i \rangle \;,
\label{eq:mA}
\end{equation}
is given in Fig.~\ref{fig:mAvsxy}. $m_{\rm B}$ is defined analogously with a sum over the 
B sublattice. A contour plot for $m_{\rm B}$ is obtained from the one for $m_{\rm A}$ by 
reflection about the $y=x$ ferromagnetic axis. In this Appendix we use the parameters 
$A=7.9$, $(T,H)=(2.820,0.149)$, and $L=6$, corresponding to the tristable region III in the 
phase diagram shown in Fig.\ \ref{Fig2_Phase_diag_L6}(d). 

\begin{figure}
\centerline{
\includegraphics[angle=0,width=.48\textwidth]{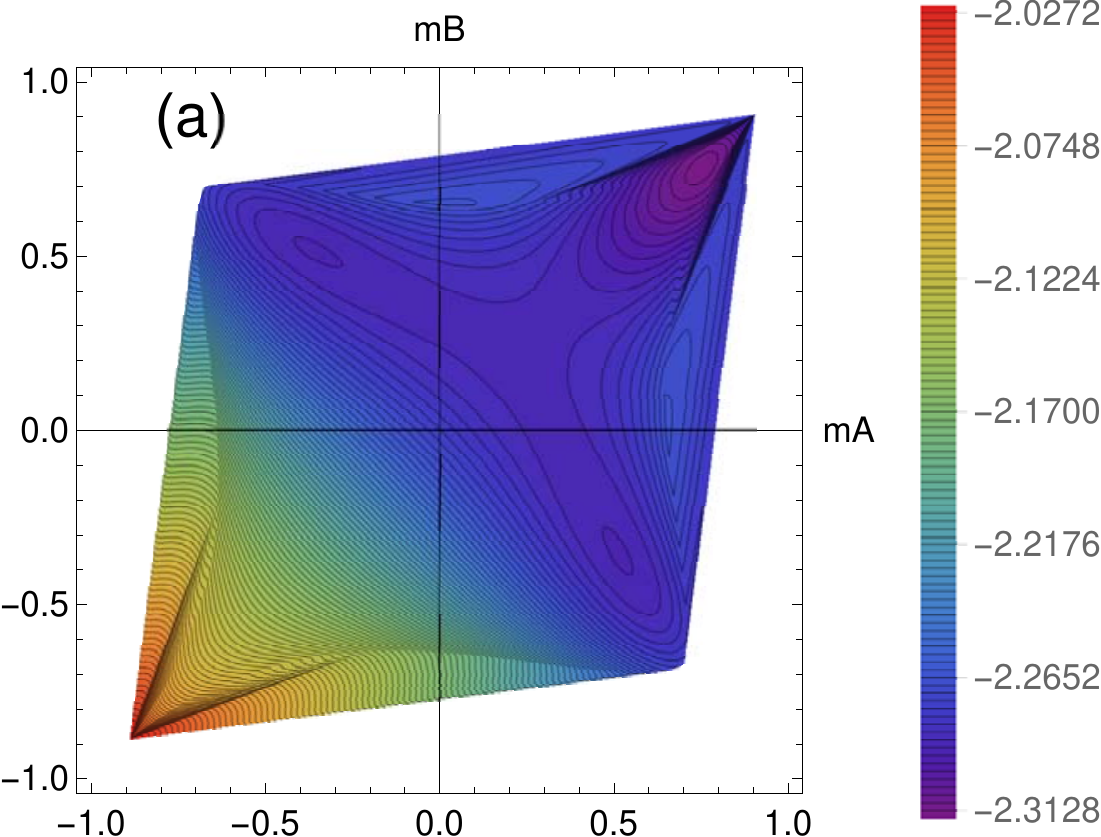}
\hspace{0.5cm} 
\includegraphics[angle=0,width=.45\textwidth]{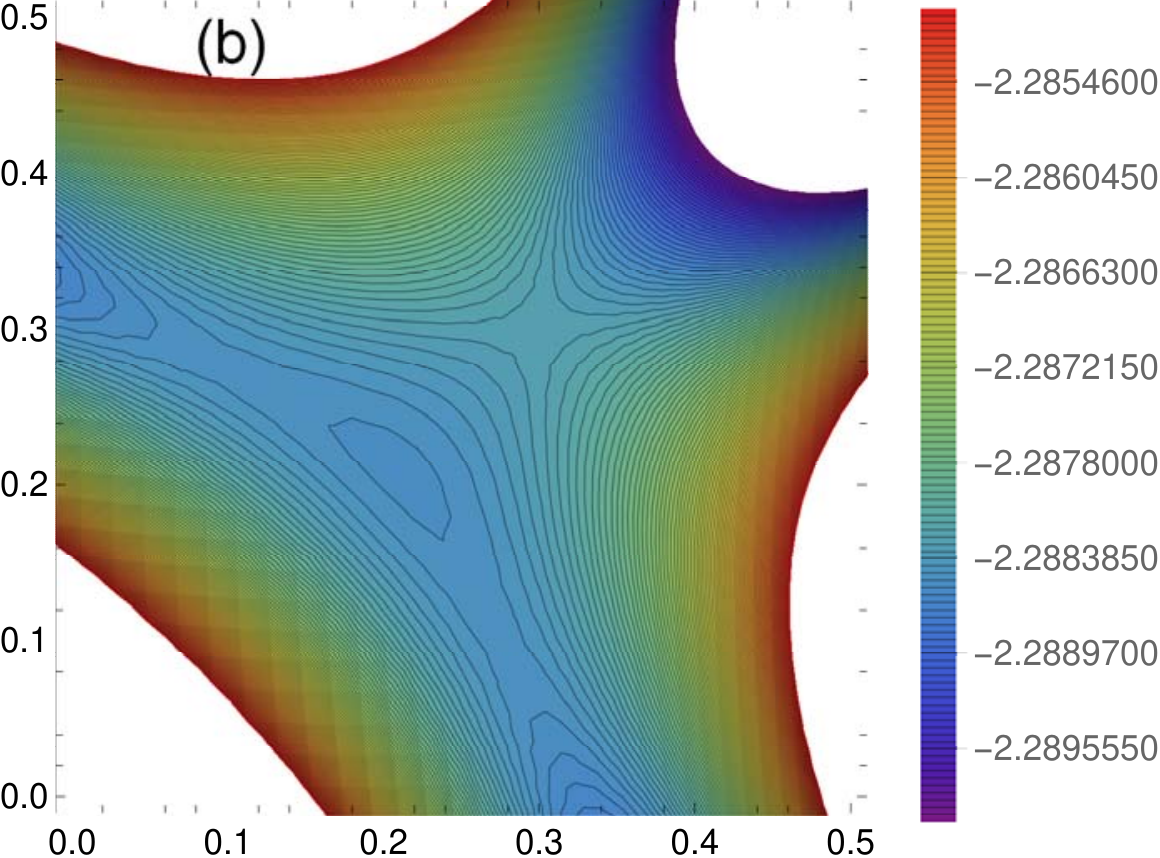}
\vspace*{0.5cm}
}
\caption{
Contour plots showing $F_{\rm v}/N$ vs  the sublattice magnetizations, 
$m_{\rm A}$ and $m_{\rm B}$, analogous to those shown vs $x$ and $y$ in 
Fig.\ \ref{Fig3_contour_regionIII}. 
(a) 
Shown over $(m_{\rm A},m_{\rm B}) \in [-1,+1] \times [-1,+1]$. 
(b) 
Shown over the restricted region, $(m_{\rm A},m_{\rm B}) \in [0,+0.5] \times [0,+0.5]$. 
The very shallow minimum corresponding to the secondary metastable D phase is clearly 
visible at this magnified scale.
 }
\label{fig:FvsmAmB}
\end{figure}

Contour plots of the per-site variational free energy $F_{\rm v}/N$ vs 
$m_{\rm A}$ and $m_{\rm B}$ 
are shown in Fig.\ \ref{fig:FvsmAmB}. The parameters are the same as in the contour plots 
of $F_{\rm v}/N$ vs $x$ and $y$ in Fig.\ \ref{Fig3_contour_regionIII}.
As seen from Fig.\ \ref{fig:mAvsxy}, varying $x$ and $y$ over the range $[-1,+1]$ does not 
cause the sublattice magnetizations to vary over this full range. As a result, the contour plot in 
Fig.\ \ref{fig:FvsmAmB}(a) covers a rhombus embedded in the $[-1,+1] \times [-1,+1]$ 
square with its long axis in the ferromagnetic direction 
$m_{\rm B} = m_{\rm A}$ and its short axis in the antiferromagnetic direction 
$m_{\rm B} = - m_{\rm A}$. 
The nonmonotonicity of $m_{\rm A}$ with respect to $y$ and of 
$m_{\rm B}$ with respect to $x$ also cause narrow, multivalued regions of $F_{\rm v}/N$ along 
the edges of the rhombus in this representation. The global minimum corresponding to the 
stable F+ phase, and the two local minima corresponding to the metastable AF phases are 
clearly seen in Fig.\ \ref{fig:FvsmAmB}(a). In the magnified plot, 
Fig.\ \ref{fig:FvsmAmB}(b), the shallow local minimum corresponding to the secondary metastable 
D phase is also seen, as well as the saddle points separating it from the metastable AF and stable 
F+ phases. 

\end{document}